\documentclass{SIP}
\usepackage{bm,amsmath,amssymb,amsfonts,amsthm,algorithm,algorithmic}
\usepackage{amssymb,float,booktabs,tikz,xcolor,multirow,threeparttable,booktabs}

\renewcommand\tabcolsep{5.0pt}
\graphicspath{{figures/}}

\begin{document}
\title[Bursty Downlink User Scheduling Using MARL]{Fairness-Oriented User Scheduling for Bursty Downlink Transmission Using Multi-Agent Reinforcement Learning}
\author[Mingqi Yuan, \textit{et al}.]{Mingqi Yuan$^{1}$, Qi Cao$^{1}$, Man-On Pun$^{1,}$$^{2}$ and Yi Chen$^{1,}$$^{2}$}
\address{\add{1}{School of Science and Engineering, The Chinese University of Hong Kong, Shenzhen, China}
	\add{2}{Shenzhen Research Institute of Big Data, Shenzhen, China}}
\corres{\name{Man-On Pun}\email{SimonPun@cuhk.edu.cn}}

\begin{abstract}
In this work, we develop practical user scheduling algorithms for downlink bursty traffic with emphasis on user fairness. In contrast to the conventional scheduling algorithms that either equally divide the transmission time slots among users or maximize some ratios without practical physical interpretations, we propose to use the \textit{5\%-tile user data rate} (5TUDR) as the metric to evaluate user fairness. Since it is difficult to directly optimize 5TUDR, we first cast the problem into the stochastic game framework and subsequently propose a \textit{Multi-Agent Reinforcement Learning} (MARL)-based algorithm to perform optimization on the resource block group (RBG) allocation in a highly computationally efficient manner. Furthermore, each MARL agent is designed to take information measured by network counters from multiple network layers (e.g. Channel Quality Indicator, Buffer size) as the input \textit{states} while the RBG allocation as \textit{action} with a carefully designed \textit{reward} function developed to maximize 5TUDR. Extensive simulation is performed to show that the proposed MARL-based scheduler can achieve fair scheduling while maintaining good average network throughput as compared to conventional schedulers.
\end{abstract}

\maketitle

\section{Introduction}
Future wireless networks have been envisaged to provide high-quality data services to a massive number of devices simultaneously \cite{dang2019human,8869705}. To accomplish this demanding goal, intensive research has been devoted to investigating how to allocate limited network resources to multiple users in an effective and yet, fair manner. In particular, user scheduling in the downlink transmission that concerns which active users should be selected and allocated network resources for transmission at the current time slot has drawn much research attention. By equally dividing the transmission time to each user, the Round Robin scheduling (RRS) can achieve the time-fairness at the cost of the network throughput \cite{hahne1991round}. In contrast, the opportunistic scheduling (OPS) was designed to allocate network resources to the advantageous users of the best instantaneous channel conditions \cite{liu2001opportunistic}. However, OPS attains the highest network throughput by scarifying those users of poor channel conditions, which incurs unfair resources allocation among users. To balance user fairness and network throughput, the proportional fairness scheduling (PFS) algorithm was proposed in the seminal work \cite{tse2001multiuser} by assigning scheduling priorities to users of the largest ratios between their instantaneous feasible data rates and historical average data rates. Recently, a parameterized PFS called Generalized PFS (GPFS) was proposed by generalizing the ratio defined in PFS with different weights  \cite{wengerter2005fairness, sang2006flexible, zhang2019comparison}. However, both PFS and GPFS only take advantages of information from one single layer, i.e. the physical  (PHY) layer. Since the network throughput is governed by protocols across multiple network layers, more comprehensive information from different network layers is required to better characterize the dynamics of the network throughput.

To cope with this problem, a machine learning (ML) approach has recently been proposed for user scheduling \cite{zhu2020toward}. In sharp contrast to the conventional model-based approach, the data-driven ML approach exploits the massive data retrieved from the network without explicitly deriving the mathematical optimization model. In \cite{comcsa2018towards}, a reinforcement learning (RL)-based algorithm was proposed to intelligently select the best scheduling algorithm from a set of pre-defined algorithms in each transmission time interval (TTI) based on the network conditions. Despite its good performance, the RL-based scheduler proposed in \cite{comcsa2018towards} suffers from high computational complexity. In \cite{xu2020buffer}, an RL-based scheduler was developed by taking into account multiple performance metrics including the estimated instantaneous data rate and the averaged data rate in constructing the input state space. In addition, an RL-enabled scheduler was also derived to optimize the network delay performance in \cite{sharma2020deep}. Recently, some pioneering attempts have been made by applying the advanced neural network structure to solve the scheduling problem. For instance, a generative adversarial network (GAN)-empowered deep distributional RL method was proposed in \cite{hua2019gan} to allocate radio resource groups in a demand-aware resource management problem. Furthermore, a pointer network architecture was adopted into the RL framework to provide flexible and scalable scheduling to cope with the network dynamics in \cite{al2020radio} and \cite{robinson2021downlink}. Finally, a knowledge-assisted deep RL was designed for scheduling to improve the training and convergence behavior in \cite{gu2021knowledge}.

However, all the aforementioned scheduling algorithms, regardless of their model-based or data-driven nature, were established upon the assumption of full-buffer traffic, i.e. all users have infinite amount of data for downlink transmission. However, all traffic in practical networks is bursty, i.e. each user's requested data volume is finite. If bursty traffic is considered, the concept of time-fairness becomes rather non-trivial. Since it takes less transmission time slots for users of a smaller amount of bursty data to finish their transmissions in general, it is reasonable to argue that equally dividing transmission time slots among all users is unfair for users requesting for more data. As a result, the concept of time-fairness becomes very subjective for the bursty traffic case. Some pioneering works on scheduling bursty traffic were reported in the literature. In \cite{li2019novel}, two novel concepts, namely the average user perceived throughput (UPT) and the user perceived throughput-cut (UPT-cut), were proposed to measure user fairness before a percentage proportional fair scheduling (PPFS) algorithm was devised to allocate more transmission time slots to users of a larger amount of bursty data. Furthermore, PPFS improves the average UPT by assigning higher priorities to users of less remaining transmission data. Recently, a hybrid downlink scheduling approach was proposed in \cite{nasralla2020hybrid} to serve bursty traffic classes of different Quality of Service (QoS) requirements. More specifically, four important flow parameters, namely the channel conditions, the packet delay, the flow queue size and the flow type, were taken into account to compute the scheduling metric using the Jain's index for each flow before assigning each resource block (RB) to the flow of the highest metric in \cite{nasralla2020hybrid}. However, the scheduling algorithms developed in both \cite{li2019novel} and \cite{nasralla2020hybrid} were designed through optimizing ratio-based metrics. Since such ratios do not have practical physical interpretations, these ratio-based algorithms cannot provide any guarantee on the actual data rate achieved by each user.

Inspired by the discussions above, we argue that rate-fairness is a more appropriate metric in designing scheduling algorithms for bursty traffic. In particular, we consider the 5\%-tile user data rate (5TUDR), i.e., the $5\%$ worst user data rate, to evaluate the user fairness. Unlike the time-fairness, the 5TUDR metric caters to inferior users without ignoring the overall network throughput. As a result, it is feasible to optimize the 5TUDR metric while implicitly maintaining a satisfactory average user data rate (AUDR). Indeed, the 5TUDR metric has been commonly employed to evaluate network performance by most wireless network operators in practice \cite{liu2001opportunistic}. However, it is challenging to use 5TUDR in practical network design due to the following reasons. First, since AUDR is governed by factors across multiple network layers, it is technically infeasible to develop a mathematical model to optimize AUDR or 5TUDR. Furthermore, 5TUDR is a long-term performance metric over multiple TTIs after a sequence of resource allocation decisions. Thus, 5TUDR cannot be directly optimized by any single-step scheduling algorithms such as PFS. In particular, if we consider allocating multiple resource block groups (RBGs) to multiple users, the optimization space grows exponentially with time, which makes the design problem analytically intractable.

To overcome these challenges, we propose to first cast the RBG allocation task as a cooperative game in which multiple RL agents collaboratively optimize a common objective function using the same reward function. After that, we devise a computationally efficient Multi-Agent Reinforcement Learning (MARL)-based scheduling algorithm to learn a low-complexity policy by decomposing the action space into multiple smaller action spaces. The main contributions of this paper are summarized as follows:

\begin{itemize}
\item We propose to model the RBG allocation process as a stochastic game by taking into account information from multiple network layers for both full-buffer and bursty downlink traffic. To maximize the user fairness without sacrificing the network throughput, we propose a novel reward function specifically designed to optimize 5TUDR;

\item Based on the proposed stochastic game, we then develop a MARL-based algorithm to optimize the scheduling policy that achieves good 5TUDR performance while maintaining a considerable AUDR.

\item Finally, extensive simulation is performed to validate the performance of the proposed MARL scheduler. The experiment results are analyzed in details to provide insights about the learned scheduling policy.
\end{itemize}

The remainder of the paper is organized as follows: Sec.~\ref{section_smpf} provides the system model and the problem formulation. Sec.~\ref{section_sgfra} elaborates the framework of the stochastic game for RBG allocation. After that, Sec.~\ref{section_marl} proposes a MARL-based algorithm for solving the stochastic game. Finally, Sec.~\ref{section_results} shows the simulation results before Sec.~\ref{section_conclusion} concludes the paper.

\underline{Notation}: Uppercase boldface and lowercase boldface letters are used to denote matrices and vectors, respectively. $\bm{I}_N$ represents the identity matrix with size $N\times N$. ${\bm A}^T$ and ${\bm A}^H$ are the transpose and conjugate transpose of ${\bm A}$, respectively. $[\bm{a}](i)$ denotes the $i$-th element of vector ${\bm a}$. In addition, $\|\bm{A}\| $ stands for the Frobenius norm of ${\bm A}$ while $|A|$ denotes the absolute value of $A$. Finally, $|\mathcal{A}|$ is the cardinality of the enclosed set $\mathcal{A}$.

\section{System Model and Problem Formulation}\label{section_smpf}
\subsection{System Model}\label{sec:sysmodel}
We consider a wireless network in which a base station (BS) schedules $K$ RBGs to multiple active user equipments (UEs) in the downlink operating in the Frequency Division Duplexing (FDD) mode. The UEs arrive at random. %arrival is modeled as a random process following the Possion distribution with an arrival rate of $\lambda$.
Upon arrival, each UE requests a finite amount of traffic data from the BS. The BS divides its frequency resources into RBGs with each RBG being allocated to at most one UE in each TTI. Furthermore, if a user is scheduled, its requested data will be transferred to a Hybrid Automatic Repeat reQuest and Retransmission (HARQ) buffer. For each transmitted package, the ACK/NACK message is fed back from the targeted UE at a fixed time interval. Upon receiving a NACK message, the BS will re-transmit the corresponding data package. A UE departs from the network immediately after all its requested data is successfully received, which effectively emulates the bursty traffic mode. In our work, some practical network mechanisms such as the Out Loop Link Adaptation (OLLA) are included in our model. The workflow of the system is shown in Fig.~\ref{fig_system} and more details about the network mechanisms considered in this work can be found in Appendix~\ref{appedix_a}.

\begin{figure}[htp]
\centering
\includegraphics[width=\linewidth]{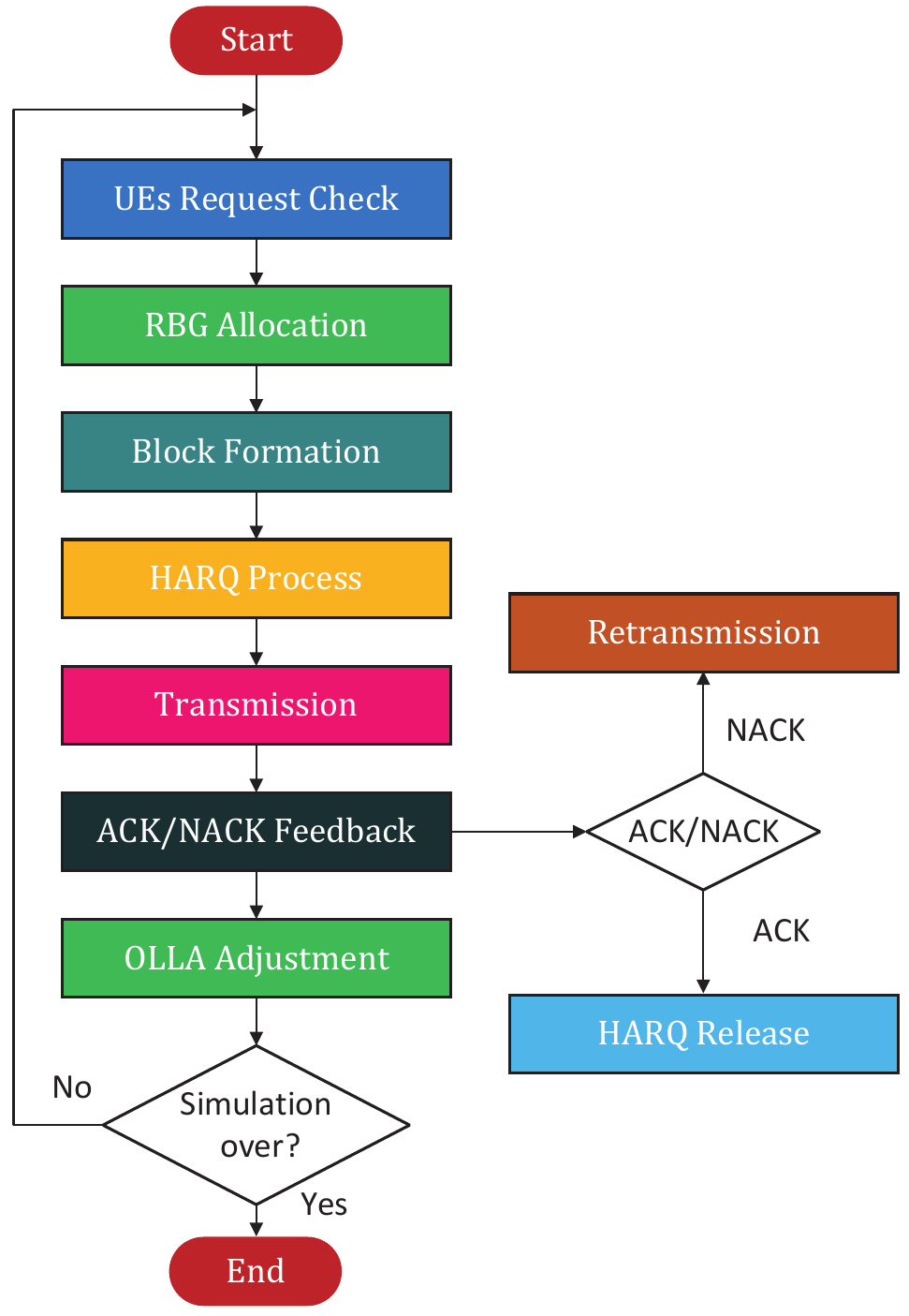}
\caption{Flowchart of the network under consideration.}\label{fig_system}
\end{figure}

\subsection{Problem Formulation}
We begin with the definitions of two key metrics, namely AUDR and 5TUDR, as well as cooperative games employed in this work.

\subsubsection{Average User Data Rate (AUDR)}
Without loss of generality, we focus on the first $T$ TTIs. We denote by $t^{n}_{a}$ and $t^{n}_{d}$ the TTI indices when the $n$-th user enters and exits the network, respectively, with $1\leq t^{n}_{a}\leq t^{n}_{d}\leq T$. Furthermore, we denote by $N_{t}$ the number of users who have arrived by the $t$-th TTI with $t\in\left[1,T\right]$. It should be emphasized that some of these $N_t$ users may have left the network by the $t$-th TTI if they finish their transmissions earlier than the $t$-th TTI. Thus, we can express the corresponding AUDR over $[1,T]$ as:
\begin{equation}
D(t) = \frac{1}{N_{t}}\sum_{n=1}^{N_{t}}\psi^{n}(t),
\end{equation}
where $\psi^{n}(t)$ is the user data rate (UDR) of the $n$-th user over $\left[t^{n}_{a},t\right]$ with $t\geq t^{n}_{a}$ and takes the following form:
\begin{equation}\label{eq:psi}
\psi^{n}(t) =\frac{\displaystyle\sum_{i=t^{n}_{a}}^{t}T^{n}[i]I^n[i]}{\min(t,t^{n}_{d})-t^{n}_{a}},
\end{equation}
where $\psi^{n}(t)=0$ if $t=t_{a}^{n}$. In Eq.~\eqref{eq:psi}, $T^{n}[i]$ is the packet size designated for the $n$-th user during the $i$-th TTI. Furthermore, $I^n[i]$ is the indicator function for the ACK/NACK feedback with ``1" indicating a successful transmission while ``0" a transmission failure. Note that the numerator in Eq.~\eqref{eq:psi} stands for the total amount of successfully transmitted data while the denominator the time duration of the $n$-th user spent in the system.

\subsubsection{$5\%$-Tile User Data Rate (5TUDR)}
In statistics, the $n$-th percentile of a set of data is defined as the value below which $n$ percent of the data falls. We denote by $\boldsymbol\Psi(t)$ the set containing $N_t\geq 1$ user data rates at time instance $t$ where
\begin{equation}
	\boldsymbol\Psi(t)=\left\{\psi^{1}(t),\ldots,\psi^{N_{t}}(t)\right\}.
\end{equation}
Without loss of generality, we assume $\psi^{i}(t)\leq \psi^{j}(t)$ where $1\leq i<j\leq N_t$. Thus,  the fifth percentile of the user data rate denoted by $\phi_{t}$ can be expressed as
\begin{equation}
\phi_{t}=P_{5\%}(\boldsymbol\Psi(t))=\psi^{z}(t),
\end{equation}
where  $P_{5\%}(\cdot)$ is the operator to find the $5\%$-tile value in the enclosed set and $z$ is given by
\begin{equation}
z=\lceil N_t\times 0.05\rceil,
\end{equation}
with $\lceil\cdot\rceil$ being the smallest integer not smaller than the enclosed number.

\subsubsection{Cooperative Games}
Stochastic games can be considered as the generalization of Markov Decision Process (MDP) with multiple agents. If all agents work together to maximize a collective return, then the stochastic game is called a cooperative game \cite{branzei2008models}. A cooperative game can be defined by a tuple $\mathcal{E}=\langle\mathcal{K},  \mathcal{S}, \mathcal{U}, P, r, \mathcal{O}, O, \gamma\rangle$ \cite{Shapley1953Stochastic}, where:
\begin{itemize}
	\item $\mathcal{K}$ is the set of agents;
	\item $\mathcal{S}$ is the global state space;
	\item $\mathcal{U}$ is the local action space;
	\item $P({\bm s}'|{\bm s},\textbf{u})$ of $\mathcal{S}\times\boldsymbol{\mathcal{U}}\times\mathcal{S}\rightarrow[0,1]$ is the transition probability, where $\mathbf{s},\mathbf{s}^{\prime}\in\mathcal{S}$ and $\textbf{u} \in \boldsymbol{\mathcal{U}}\equiv\mathcal{U}^{|\mathcal{K}|}$;
	\item $r({\bm s},\textbf{u})$ of $\mathcal{S}\times\boldsymbol{\mathcal{U}}\rightarrow\mathbb{R}$: the reward function that evaluates the reward for a given state-action pair;
	\item $\mathcal{O}$ is the local observation space with ${\bm o}_{k}\in\mathcal{O}$;
	\item $O({\bm s},k)$: $\mathcal{S}\times\mathcal{K}\rightarrow\mathcal{O}$ is the observation function;
	\item $\gamma\in[0,1]$ is a discount factor.
\end{itemize}

Using the definition above, a policy can be defined as a mapping from observations to actions \cite{sutton2018reinforcement}. Mathematically, a policy denoted by $\boldsymbol{\pi}({\bm s})$ comprises the suggested actions that the agents should take for every possible state ${\bm s}\in\mathcal{S}$. For notational simplicity, the input notation ${\bm s}$ is omitted and a policy is simply represented as $\boldsymbol{\pi}$ in the sequel.

Finally, we are ready to formulate our problem using the three definitions above. Given a network described in Sec.~\ref{section_smpf}.\ref{sec:sysmodel}, our goal is to establish an optimal scheduling policy denoted by $\boldsymbol{\pi}$ that maximizes the rate-fairness metric 5TUDR over the time interval $\left[1,T\right]$. Specifically, the optimization problem can be written as
\begin{align}\label{eq:maxoo}
\underset{\boldsymbol{\pi}}{\rm argmax} & \quad\phi_{T}\left(\boldsymbol{\pi}\right)\tag{OP1}\\
{\rm s.t.} &\quad C_1: D(T) \geq \kappa, \nonumber\\
%&\quad C_2:P(N_{t})=\frac{\lambda^{N_{t}}}{N_{t}!}e^{-\lambda},\quad\forall t\in[1,T]\nonumber
&\quad C_2:P(N_{t}-N_{t-1}=n)= f(n)\quad\forall t\in[1,T],\nonumber
\end{align}
where $\kappa$ in Constraint $C_1$ is the minimum required AUDR while $C_2$ defines that the probability of having $n$ new arriving users follows some distribution $f(n)$. For example, when the Poisson distribution with an average arrival rate of $\lambda$ is considered, $f(n)=\frac{\lambda^{n}}{n!}e^{-\lambda}$.

Clearly, it is non-trivial to find the optimal policy $\boldsymbol{\pi}^*$ by directly solving (OP1) due to two reasons. First, it is analytically difficult to manipulate $\phi_{T}\left(\boldsymbol{\pi}\right)$ as it does not have a closed-form expression. Second, the optimization space of (OP1) grows polynomially with the number of active users, and exponentially with the number of RBGs under consideration as well as the time duration. In the sequel, we will first cast (OP1) into a stochastic game framework before a MARL-based scheduling algorithm is developed to exploit data from multiple network layers.

\begin{figure*}[htp]
\centering
\includegraphics[width=0.9\linewidth]{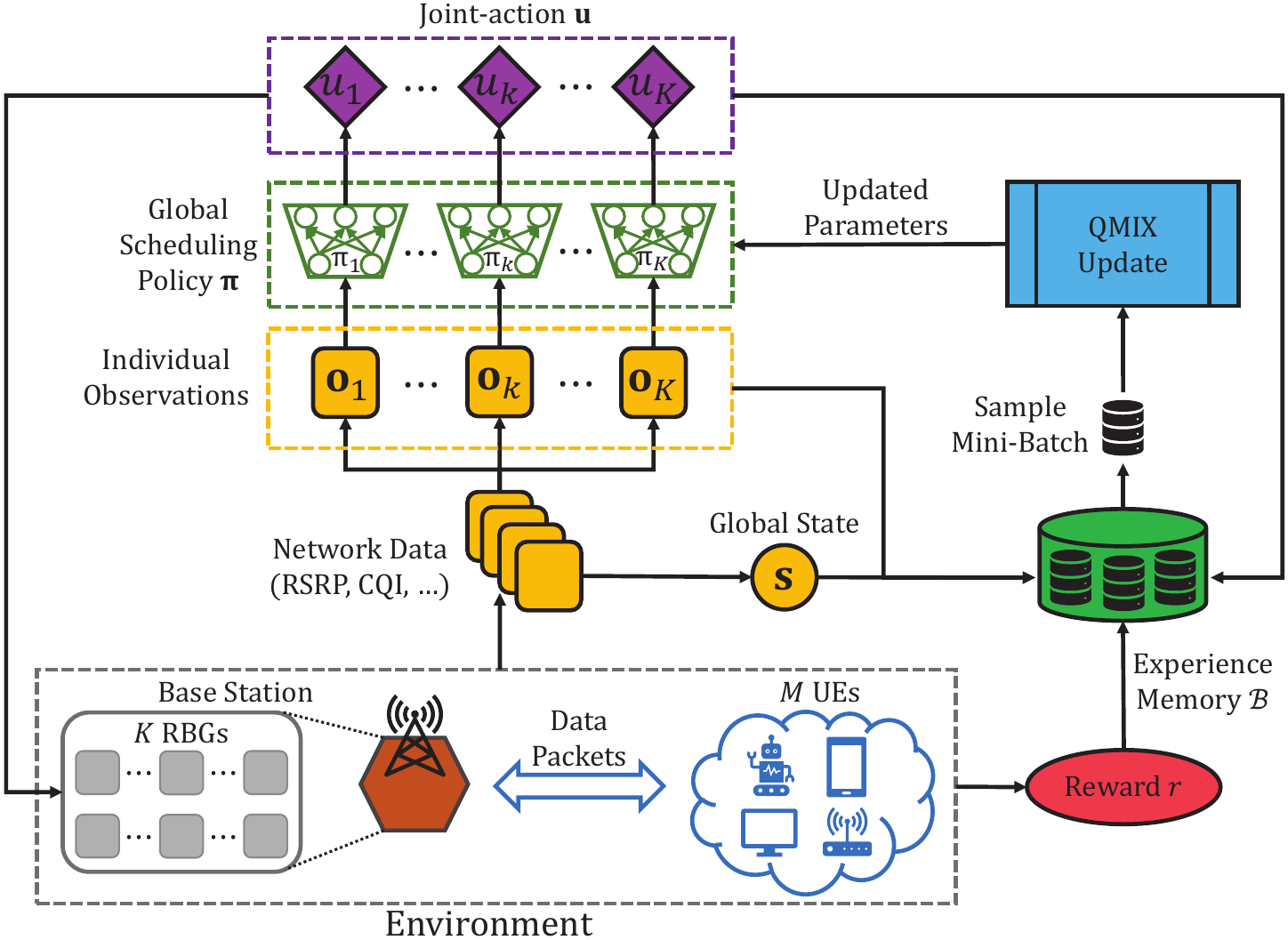}
\caption{The overview of the cooperative game of the RBG allocation.}
\label{overview}
\end{figure*}

\section{Stochastic Game Framework for RBG Allocation}\label{section_sgfra}
In this paper, we model the RBG allocation as a partially-observable cooperative game. We arrange $K$ agents denoted by $\mathcal K=\{1,\ldots,K\}$ to allocate the $K$ RBGs separately. The $k$-th agent for $k\in \mathcal K$ learns an independent scheduling policy $\pi_{k}$ to decide the allocation of the $k$-th RBG based on its individual observation. The joint policy is denoted by $\boldsymbol{\pi}=\left\{\pi_1,\pi_2,\cdots,\pi_K \right\}$. %We denote by $\mathbf{u}=(u_{1},\cdots,u_{K})$ the joint action of all $K$ agents.
%n the following discussions, we will concentrate on the scheduling decision for the $t$-th TTI. For simplicity of presentation, we will ignore the time index in the notations in this section. In this game, we arrange $K$ agents to allocate a set of $K$ RBGs independently with each agent using an independent scheduling policy $\pi_{k}$ whose output is the action $u_{k}$ representing the allocation decision of the $k$-th agent. We denote by $\mathbf{u}=(u_{1},\cdots,u_{K})$ the joint action of all $K$ agents.

Fig.~\ref{overview} illustrates the overview of the cooperative game for the task of RBG allocation. At every TTI, each of the $K$ agents first observes the environment and gathers network data to form its local observation $\mathbf{o}_{k}\in\mathcal{O}$ before choosing its action. It should be emphasized that given the problem complexity, we propose not to use the global state $\mathbf{s}\in \mathcal{S}$ that is the collection of all local observations $\left\{\mathbf{o}_k\right\}_{k\in\mathcal K}$ in this decision-making stage. %Instead, $\bm{s}$ will be used to update the scheduling policy in Sec.~\ref{section_marl}.
Upon receiving the decisions $\mathbf u = (u_1,\ldots,u_K)$, the BS performs the corresponding transmission as specified by the agents. After that, the reward function will generate a collective reward for all agents to evaluate the overall allocation performance. Finally, the observations, actions and rewards will be stored in the experience memory for the scheduling policy update. The details about the policy update will be elaborated in Sec.~\ref{section_marl}. Next, we will explain the states, the actions and the reward function employed in the cooperative RBG allocation game.

\subsection{Cross-layer Observation and Adaptive Action}\label{sec:crossinfo}
For a traditional model-based RBG scheduler, it can only utilize information from one single network layer due to the prohibitively high complexity incurred in modeling information from multiple layers. In this paper, we propose to better characterize the network status by efficiently exploiting information from multiple network layers.

\begin{table}[H]
\caption{Network data of observations}\label{tb:observations}
\centering
\begin{tabular}{llc}
	\toprule[1.5pt]
	Counter Name            & Layer   & Dimension \\ \midrule[1.5pt]
	RSRP                    & PHY     & 1         \\
	Average of RB CQI       & PHY     & 1 		  \\
	Buffer size             & MAC     & 1         \\
	Scheduled frequency     & MAC     & 1         \\
	OLLA offset         	& MAC     & 1         \\
	Historical user data rate (HUDR)  & MAC     & 1         \\ \bottomrule[1.5pt]
\end{tabular}
\end{table}

Table~\ref{tb:observations} shows the network counters selected in our design as well as the layer that each counter belongs to. For instance, Reference Signal Receiving Power (RSRP) is selected to represent the longer-term wireless signal strength while Channel Quality Indicator (CQI) is employed to reflect the instantaneous channel quality corresponding to the channel's signal-to-noise ratio (SNR). Note that the CQI value for the same user varies in different RBGs due to the frequency-selective channels commonly encountered by the broadband wireless communication systems. Furthermore, we take into account the buffer size that stands for the amount of remaining data to be transferred to an active user as well as each user's scheduled frequency that keeps track of how often each user has been scheduled for transmission. Finally, the historical user data rate (HUDR) and the OLLA offset are also included in our observations, where the HUDR is the UDR up to the last TTI. As indicated in Table~\ref{tb:observations}, all these network parameters are taken from either the physical (PHY) or the multiple access (MAC) layers. It should be emphasized that more counters can be included in our observation in a straightforward manner at the cost of higher computational complexity.

At each TTI, each agent forms its observation including the counters listed in Table~\ref{tb:observations}. We denote by $C$ and $M_t$ the total number of network counters under consideration and the number of active users at the current TTI, respectively. We can then construct an $M_t\times C$ feature matrix $\mathbf{O}_k$ for each $k\in \mathcal K$ based on their individual observations as follows:
\begin{equation}
\mathbf{O}_k=
\begin{pmatrix}
\tilde o_k^{1,1} &\cdots & \tilde o_k^{1,C}\\
\vdots &\ddots & \vdots\\
\tilde o_k^{M_t,1} &\cdots &\tilde o_k^{M_t,C}
\end{pmatrix},
\end{equation}
where $\tilde o_k^{m,c}$ stands for the $c$-th counter value over the $k$-th RBG observed by the $m$-th user with $m=1,2,\ldots, M_t$ and $c=1,2\ldots,C$. For instance, assuming that CQI is labeled as the first counter, then $\tilde o_k^{m,1}$ represents the CQI of the $m$-th user over the $k$-th RBG.

To fully exploit the features of the cross-layer information, we propose to use the deep neural network (DNN) of fixed input and output dimension to represent the individual scheduling policy. However, for \textit{bursty traffic}, the dimension of the feature matrix $\mathbf{O}_k$ varies with $M_t$ over time, which makes it incompatible with our designed DNN architecture. To circumvent this problem, we propose to apply the following vectorization operation on $\mathbf{O}_k$:
\begin{equation}\label{eq:ok}
\mathbf{o}_k=\text{Vec}\left(\mathbf{O}_{k}^T\mathbf{O}_k\right),
\end{equation}
where the operator $\text{Vec}(\cdot)$ converts the enclosed matrix into a single vector row by row. Note that the resulting vector $\mathbf{o}_k$ has a fixed length of $C^2$. It is worth noting that Eq.~\eqref{eq:ok} is one of the many possible encoding scheme to generate a vector of a fixed length from $\mathbf{O}_{k}$.

Capitalizing on $\mathbf{o}_k$ of a fixed length as the input, the DNN generates preference values to all $M_t$ active users for each RBG before assigning the RBG to the best user. However, recalling that the number of active users $M_t$ also varies with time, i.e. the output of the DNN has a varying length. In order to have a fixed size of the DNN output, we propose to pre-define a maximum allowable number of active users denoted by $M_{\max}$. Therefore, at each decision step, $M_{\max}$ preference values are generated, but only the first $M_t$ values are used to determine the RBG allocation.

\subsection{Reward Function}
Next, we devise a novel reward function to evaluate the performance of each agent. Since the design goal is to maximize 5TUDR while maintaining a considerable AUDR, the desired reward function should be able to evaluate the contribution of each scheduling decision towards the final 5TUDR and AUDR. A trivial 5TUDR-maximizing reward function can be designed to maximize the increment in 5TUDR between two consecutive TTIs. Unfortunately, the performance of such reward functions has been found very poor with bursty traffic.

\begin{figure}[htp]
\centering
\includegraphics[width=\linewidth]{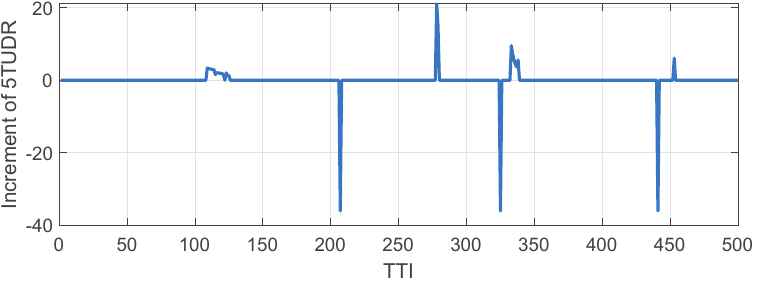}
\caption{An example of the increment (decrement when negative) of 5TUDR between two consecutive TTIs over time.}
\label{fig_r1}
\end{figure}

Fig.~\ref{fig_r1} shows the increment of 5TUDR as a function of time in TTI. Inspection of Fig.~\ref{fig_r1} reveals that the increment of 5TUDR is always $0$ with occasional sharp increases and decreases. Careful investigation on those sharp spikes indicates that those sharp decreases were caused by the arrival of new users of $\psi(t)=0$. In contrast, once a new user was scheduled, its user data rate suddenly became non-zero, which caused a sharp increase in the increment of 5TUDR. Therefore, using the increment of 5TUDR as the reward function is not appropriate for bursty traffic as the increment is a random event due to the arrival or departure of users.

\begin{figure}[htp]
\centering
\includegraphics[width=\linewidth]{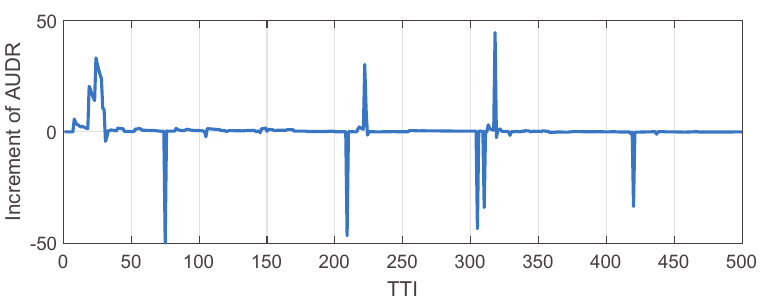}
\caption{An example of the increment (decrement when negative) of AUDR between two consecutive TTIs over time.}
\label{fig_r2}
\end{figure}

Alternatively, we can consider employing the increment of AUDR as the reward function. However, the increment of AUDR is also heavily influenced by the arrival or departure of users. Fig.~\ref{fig_r2} illustrates an example of the increment of AUDR between two consecutive TTIs. In Fig.~\ref{fig_r2}, the sharp decreases of AUDR were caused by the arrival of new users as their contribution of zero data rates to AUDR. In contrast, the sharp increases occurred when the agents allocated RBGs to users of advantageous channel conditions and large data packets were transmitted in a short period of time. Clearly, the increment of AUDR is not an appropriate reward function to maximize 5TUDR either.

To strike the balance between 5TUDR and AUDR, we propose the following reward function by exploiting the changes of the {\em sum} of UDR over two consecutive TTI's, respectively.
\begin{eqnarray}\label{eq:reward}
r(t)&=&h\left(\sum_{n=1}^{M_{t}}\psi^{n}(t)-\sum_{n=1}^{M_{t-1}}\psi^{n}(t-1)\right)\\\nonumber
&&-\exp\{-\mathcal{G}(\Delta\boldsymbol{\Psi}(t))\},
\end{eqnarray}
where $h(\cdot)$ is the sigmoid function and $\Delta\boldsymbol{\Psi}(t)$ stands for the UDR change of each active user given by
\begin{equation}
\Delta\boldsymbol{\Psi}(t)=\big\{\Delta\psi^{1}(t),\cdots,\Delta\psi^{M_{t}}(t)\big\},
\end{equation}
with $\Delta\psi^{n}_{t}=\psi^{n}(t)-\psi^{n}(t-1)$ for $n=1,2,\ldots,M_t$. In addition, $\mathcal{G}(\cdot)\in[0,1]$ is the G's fairness index that measures the level of variations of the elements of the enclosed set \cite{jain1984quantitative}. The G's fairness index takes value zero when there is a zero element. For instance, the Jain's fairness index proposed in \cite{jain1984quantitative} is employed in our experiments. Mathematically, the Jain's fairness index of a set of values $\left\{x_i\right\}$, where $i=1,2,\cdots,\ell$, is given by :
\begin{equation}\label{eq:jain}
	\mathcal{G}\left(x_{1}, x_{2}, \ldots, x_{\ell}\right)=\frac{\left(\displaystyle\sum_{i=1}^{\ell} x_{i}\right)^{2}}{\ell \cdot \displaystyle\sum_{i=1}^{\ell} x_{i}^{2}}.
\end{equation}
It can be shown from Eq.~\eqref{eq:jain} that the best case, i.e. $x_1=x_2=\cdots=x_\ell$, has a Jain's fairness index of $1$. Furthermore, the worst case corresponding to the scenario that one of the values is much larger than the other $\ell-1$ values has a Jain's fairness index of $\frac{1}{\ell}$ \cite{jain1984quantitative}. It is worth mentioning that any G's fairness index functions that penalize uneven UDR changes across active users can be employed in Eq.~\eqref{eq:reward}. Furthermore, in sharp contrast to the ratios used in the conventional PFS schemes that do not directly correspond to the 5TUDR performance, the G's fairness index in Eq.~\eqref{eq:reward} is designed to explicitly make all active users' UDR uniform while maximizing the sum of UDR over all active users.

The following features of the proposed reward function in Eq.~\eqref{eq:reward} deserve further discussion. First, the reward function is conditioned upon the sum of all user data rates, which makes the reward function insensitive to the time-varying number of total arrived users. In particular, when new users of zero UDR enter the system, the reward function is not affected. Second, since the reward function effectively maximizes the increment in the user data sum-rate, the resulting scheduling policy drives all agents to effectively allocate RBGs. Finally, the G's fairness index is introduced as the regularization term to discourage agents to be in favor of a few users. As a result, more users will be given transmission opportunities, which effectively improves 5TUDR.

With the proposed reward function, we are ready to propose the following alternative optimization problem:
\begin{align}\label{eq:maxsumrate}
\underset{\boldsymbol{\pi}}{\rm argmax} \quad& \mathbb{E}_{\boldsymbol{\pi}}\bigg[\sum_{t=1}^{T}\gamma^{t}r(t)\bigg] \tag{OP2}\\
\text{s.t.} \quad &C_1: D(T) \geq \kappa\nonumber\\
&C_2:P(N_{t}-N_{t-1}=n)= f(n)\quad\forall t\in[1,T],\nonumber%C_2:P(N_{t})=\frac{\lambda^{N_{t}}}{N_{t}!}e^{-\lambda}\quad\forall t\in[1,T]\nonumber
\end{align}
where $\mathbb{E}_{\boldsymbol{\pi}}[\cdot]$ denotes the expected value of a random variable given that the agents follow policy $\boldsymbol{\pi}$. It is worth noting that  \eqref{eq:maxsumrate} is designed to optimize the total user sum rate while penalizing any large discrepancy among user data rates. Thus, \eqref{eq:maxsumrate} {\em implicitly} optimizes 5TUDR without requiring the explicit closed-form expression of 5TUDR. Extensive simulation results will be shown in Sec.~\ref{section_results} to demonstrate that the solution to \eqref{eq:maxsumrate} strikes a good balance between AUDR and 5TUDR as compared to GPFS.

In particular, \eqref{eq:maxsumrate} indicates that the objective of this cooperative game is to find a policy $\boldsymbol{\pi}$ that maximizes the long-term return. All agents are allowed to explore diverse polices as long as \eqref{eq:maxsumrate} is maximized. For such a cooperative game problem, the Nash Equilibrium (NE) is commonly employed to describe the solution \cite{myerson2013game}. Unfortunately, it is non-trivial to find NE in many practical scenarios. To cope with this problem, a learning algorithm based on MARL is proposed in the next section.

\section{MARL-Based Algorithm}\label{section_marl}

\begin{figure*}[h]
\centering
\includegraphics[width=0.85\linewidth]{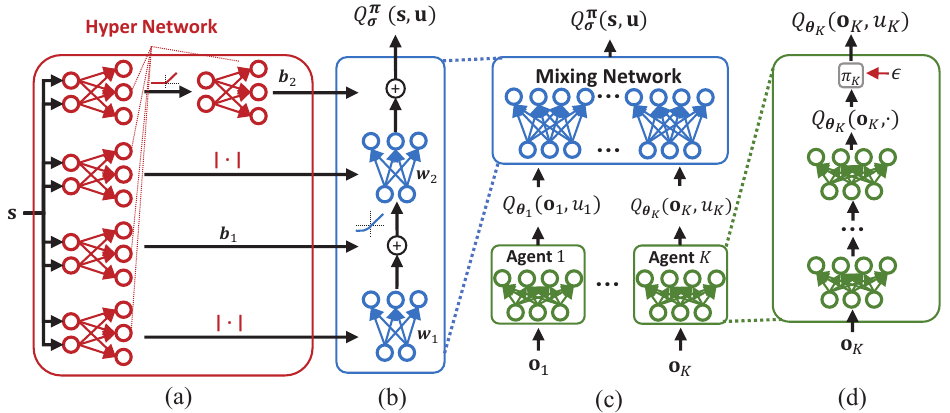}
\caption{(a) The hyper-network that produces the weights and biases for mixing network layers shown in blue. (b) Mixing network structure. (c) The overall QMIX architecture. (d) Agent network structure.}
\label{fig:qmix}
\end{figure*}

In this section, we will first review some basics about MARL before introducing a popular MARL algorithm called QMIX \cite{rashid2018qmix}.

\subsection{MARL Background}
MARL learns from the interactions with the environment and subsequently, updates its policy to maximize the long-term return. To evaluate the policy, the action-value function is leveraged to characterize the expected return. More specifically, the expected return takes the following form if the agents execute the joint-policy $\boldsymbol{\pi}$ after taking joint-action $\mathbf{u}$ for the global state $\mathbf{s}$:

\begin{equation}
Q^{\boldsymbol{\pi}}(\mathbf{s},\mathbf{u}) =\mathbb{E}_{\boldsymbol{\pi}} \bigg[ \sum_{t=0}^{\infty}\gamma^{t}r_{t} |\mathbf{s}_{0}=\mathbf{s},\mathbf{u}_{0}=\mathbf{u}\bigg].
\end{equation}

The Bellman optimal theorem indicates that the optimal policy $\boldsymbol{\pi}^{*}$ satisfies:
\begin{equation}\label{eq:bellman_optimal}
Q^{*}(\mathbf{s},\mathbf{u})=Q^{\boldsymbol{\pi}^{*}}(\mathbf{s},\mathbf{u})=\sup_{\boldsymbol{\pi}\in\Pi}\{Q^{\boldsymbol{\pi}}(\mathbf{s},\mathbf{u})\},
\end{equation}
where $\Pi$ is the set of stationary policies. Moreover, the $Q^{*}(\mathbf{s},\mathbf{u})$ obeys the Bellman optimality equation \cite{bellman1966dynamic}:
\begin{equation}\label{eq:boe}
Q^{*}(\mathbf{s},\mathbf{u})=\mathbb{E}_{s'\sim P(\cdot|\mathbf s,\mathbf u)}\bigg[r(\mathbf{s},\mathbf{u})+\gamma \max_{\mathbf{u}'\in\boldsymbol{\mathcal{U}}}\{Q^{*}(\mathbf{s}',\mathbf{u}')\}\bigg].
\end{equation}
%	Once obtaining $Q^{*}(\mathbf{s},\mathbf{u})$, the optimal policy can be derived directly by
%	\begin{equation}
%	\boldsymbol{\pi}^*(\mathbf s) = \underset{\mathbf{u}\in\boldsymbol{\mathcal{U}}}{\text{argmax}}  \ Q^{*}(\mathbf{s},\mathbf{u}).
%	\end{equation}
Based on Eq.~\eqref{eq:boe}, we can estimate the optimal $Q^{*}(\mathbf{s},\mathbf{u})$ through the following iterative method:
\begin{equation}
Q_{t+1}(\mathbf{s},\mathbf{u})=\mathbb{E}_{\mathbf s'\sim P(\cdot|\mathbf s,\mathbf u)}\bigg[r(\mathbf{s},\mathbf{u})+\gamma \max_{\mathbf{u}'\in\boldsymbol{\mathcal{U}}}\{Q_{t}(\mathbf{s}',\mathbf{u}')\}\bigg].
\end{equation}
It has been proved that such an iterative method converges to the optimal action-value function, that is, $Q_{t}\rightarrow Q^{*}$ as $t\rightarrow\infty$ \cite{watkins1992q}. In practice, we usually use the function approximation to estimate the action values to overcome the high-dimensional state space. Denote by $Q^{\boldsymbol{\pi}}_{\bm \sigma}(\mathbf{s},\mathbf{u})$ the action-value function with parameters $\bm \sigma$, the optimization objective can be transformed to find $\bm \sigma$ such that $Q^{\boldsymbol{\pi}}_{\bm \sigma}(\mathbf{s},\mathbf{u})\approx Q^{*}(\mathbf{s},\mathbf{u})$. The optimal parameters $\bm \sigma$ can be trained by minimizing the following temporal difference error \cite{mnih2013playing}:
\begin{equation}\label{eq:dqnloss}
\mathcal{L}(\bm \sigma)=\bigg[ r(\mathbf{s},\mathbf{u}) + \gamma\max_{\mathbf{u}'}\{Q^{\boldsymbol{\pi}}_{\bm \sigma}(\mathbf{s}',\mathbf{u}')\}-Q^{\boldsymbol{\pi}}_{\bm \sigma}(\mathbf{s},\mathbf{u}) \bigg]^{2}.
\end{equation}

\subsection{QMIX}
In the multi-agent learning problem, it is not straightforward to learn the action-value function  $Q^{\boldsymbol{\pi}}_{\bm \sigma}(\mathbf{s},\mathbf{u})$ due to the exponentially growing dimension of the joint-action space. To address this problem, the QMIX algorithm was developed to learn the joint-action value function through value decomposition \cite{sunehag2018value}. QMIX is a model-free, value-based, off-policy, parallel execution but centralized training algorithm designed for cooperative tasks. QMIX employs the DNN-based function approximation approach to estimate the action-value function \cite{rashid2018qmix}.

Fig.~\ref{fig:qmix} illustrates the architecture of QMIX. As shown in Fig.~\ref{fig:qmix}, QMIX allows each agent $k$ to maintain an individual action-value function $Q_{{\bm \theta}_{k}}(\mathbf{o}_k,u_k)$ that is conditioned upon each individual observation $\mathbf{o}_k$ and local action $u_k$, where ${\bm \theta}_{k}$ is the parameters of the DNN approximator. In every step, each agent $k$ accepts an individual observation $\mathbf{o}_k$ before generating the corresponding action values. After that, the action is determined using the $\epsilon$-greedy principle with the probability of choosing the action $u_k$ given by \cite{wunder2010classes}:
\begin{equation}
P(u_k|\mathbf{o}_k)=
\begin{cases}
1-\epsilon + \frac{\epsilon}{|\mathcal{U}|}, & u_k=\underset{u_k}{\text{argmax}}\{Q_{{\bm \theta}_{k}}(\mathbf{o}_k,u_k)\} \\
\frac{\epsilon}{|\mathcal{U}|}, & u_k\neq\underset{u_k}{\text{argmax}}\{Q_{{\bm \theta}_{k}}(\mathbf{o}_k,u_k)\}
\end{cases},
\end{equation}
where $|\mathcal{U}|$ is the cardinality of the action space.
This policy requires each agent to randomly select an action from the action space with a probability of $\epsilon$, or follow the action with the maximum action-value with a probability of $1-\epsilon$. This policy encourages abundant explorations by assigning a non-zero probability to explore all possible state-action pairs during training.

To guarantee the consistency between $Q^{\boldsymbol{\pi}}_{\bm \sigma}$ and $Q_{{\bm \theta}_{k}}$, we only need to ensure that the global argmax operation applied on the joint $Q^{\boldsymbol{\pi}}_{\bm \sigma}(\mathbf{s},\mathbf{u})$ generates the same result as a set of argmax operation applied on each $Q_{{\bm \theta}_{k}}$:

\begin{equation}
\underset{\mathbf{u}}{\text{argmax}} \{ Q^{\boldsymbol{\pi}}_{\bm \sigma}(\mathbf{s},\mathbf{u}) \}=
\begin{pmatrix}
\underset{u_{_1}}{\text{argmax}}\{Q_{{\bm \theta}_{1}}(\mathbf{o}_{1}, u_{_1})\} \\
\vdots \\
\underset{u_{_K}}{\text{argmax}}\{Q_{{\bm \theta}_{K}}(\mathbf{o}_{K}, u_{_K})\}
\end{pmatrix}.
\end{equation}

Mathematically, this imposes a monotonicity constraint on the relationship between $Q^{\boldsymbol{\pi}}_{\bm \sigma}(\mathbf{s},\mathbf{u})$ and each $Q_{k}$:
\begin{equation}\label{eq:monotonicity}
\frac{\partial Q^{\boldsymbol{\pi}}_{\bm \sigma}(\mathbf{s},\mathbf{u})}{\partial Q_{{\bm \theta}_{k}}} \geq 0, \forall k.
\end{equation}

%If the centralized training is performed,
To enforce Eq.~\eqref{eq:monotonicity}, QMIX designs a mixing network as shown in Fig.~\ref{fig:qmix}. The mixing network is a two-layer neural network that takes the agent network outputs $Q_{\bm \theta_k}$ for $k=1,\ldots,K$ as input, and mixes them to generate the joint action-value $Q^{\boldsymbol{\pi}}_{\bm \sigma}$. In particular, the parameters $\bm \sigma$ of the mixing network are produced by the hyper-network, denoted by $\mathcal{H}_{\bm \delta}$ with parameters $\bm \delta$. With reference to Fig.~\ref{fig:qmix}, we denote by $\mathbf{w}_1,\mathbf{w}_2, \mathbf{b}_1$ and $\mathbf{b}_2$ the weights and biases of the mixing network, respectively. Given a state $\mathbf{s}$, we have% At the $t$-th TTI, the hyper-network accepts the global state $\mathbf{s}(t)$ and generates weights and biases for the mixing network:
\begin{equation}
\bm \sigma = (\mathbf{w}_1,\mathbf{w}_2, \mathbf{b}_1,\mathbf{b}_2) = \mathcal{H}_{\bm \delta}(\mathbf{s}).
\end{equation}
We set $\bm \sigma$ to be positive in order to satisfy Eq.~\eqref{eq:monotonicity}. Let $\mathbf{Q}=(Q_{{\bm \theta}_{1}},...,Q_{{\bm \theta}_{K}})$, then $Q^{\boldsymbol{\pi}}_{\bm \sigma}(\mathbf{s},\mathbf{u})$ is computed as follows:
\begin{equation}\label{eq:qsigma}
Q^{\boldsymbol{\pi}}_{\bm \sigma}(\mathbf{s},\mathbf{u}) = \mathbf{w}_2\cdot g(\mathbf{w}_1\mathbf{Q}+\mathbf{b}_1) + \mathbf{b}_2,
%	&= \mathbf{w}_2\cdot g(\mathbf{y}_{1})+\mathbf{b}_2,
\end{equation}
where
\begin{equation}\label{eq:gx}
g(x)=\begin{cases}
\begin{array}{ll}
\alpha(e^{x}-1) & x<0 \\
x, & x \geq 0
\end{array}
\end{cases},
\end{equation}
is the standard ELU activation function and applied component-wise with $\alpha>0$.

Taking the derivative of $Q^{\boldsymbol{\pi}}_{\bm \sigma}(\mathbf{s},\mathbf{u})$ with respect to $ \mathbf{Q}$, we have
\begin{equation}
\begin{aligned}
\frac{\partial Q^{\boldsymbol{\pi}}_{\bm \sigma}(\mathbf{s},\mathbf{u})}{\partial \mathbf{Q}}	%&= \frac{\partial Q_{\bm \sigma}^{\boldsymbol{\pi}}(\mathbf{s},\mathbf{u})}{\partial g(\mathbf{y})} \cdot \frac{\partial g(\mathbf{y})}{\partial \mathbf{y}} \cdot \frac{\partial \mathbf{y}}{\partial \mathbf{Q}}, \\
=
\mathbf{w}_2 \cdot g'(\mathbf{w}_1\mathbf{Q}+\mathbf{b}_1) \cdot \mathbf{w}_1\geq 0.\\
\end{aligned}
\end{equation}
The last inequality holds since $\mathbf{w}_1$ and $\mathbf{w}_2$ are non-negative and
%	Recalling Eq.~\eqref{eq:gx}, we have
\begin{equation}
g'(x) =
\begin{cases}
\begin{array}{ll}
\alpha e^{x}& x<0 \\
1& x \geq 0.
\end{array}
\end{cases}.
\end{equation}

Finally, QMIX is trained by minimizing the following loss function:
\begin{equation}\label{eq:qmixloss}
\mathcal{L}(\boldsymbol{\bm \theta},{\bm \delta},{\bm \sigma})=\big[ r(\mathbf{s},\mathbf{u}) + \gamma\max_{\mathbf{u}'}Q^{\boldsymbol{\pi}}_{\bm \sigma}(\mathbf{s}',\mathbf{u}')-Q^{\boldsymbol{\pi}}_{\bm \sigma}(\mathbf{s},\mathbf{u}) \big]^{2},
\end{equation}
where $\boldsymbol{\bm \theta}$ contains the DNN parameters of all agents. Note that Eq.~\eqref{eq:qmixloss} takes the similar form as Eq.~\eqref{eq:dqnloss} for the update operations. Finally, the QMIX-based  algorithm updates the scheduling policy as summarized in Algorithm~\ref{alg_qmix_dis}.

\begin{algorithm}[htp]
	\caption{The QMIX algorithm for User Scheduling}
	\begin{algorithmic}[1]\label{alg_qmix_dis}
		\STATE Initialize $K$ agent networks $\left\{Q_{{\bm \theta}_{1}}, \dots , Q_{{\bm \theta}_{K}}\right\}$ with DNN parameters $\boldsymbol{\bm \theta}=\left\{{\bm \theta}_{1},\dots,{\bm \theta}_{K}\right\}$;
		
		\STATE Initialize the hyper-network $\mathcal{H}_{\bm \delta}$ and the mixing network $\mathcal{M}_{\bm \sigma}$ with parameters $\bm \delta$ and $\bm \sigma$;
		\STATE Set a replay buffer $\mathcal{B}$, a learning rate $\tau$, a discount factor $\gamma$ and an exploration rate $\epsilon$;		
		
		\FOR{$t=1,\dots,T$}
		\STATE Each agent $k$ makes its local action $u_{k}(t)$ based on its local  observation $\mathbf{o}_{k}(t)$ using the $\epsilon$-greedy policy before observing a new local state $\mathbf{o}_{k}(t+1)$;
		\STATE Execute the joint-action $\mathbf{u}(t)$ and evaluate the team reward $r(t)$ before collecting a new global state $\mathbf{s}(t+1)$;
		\STATE Store the following transition in $\mathcal{B}$ for $k\in\{1,\dots,K\}$:
		\begin{equation}\nonumber
			\bigl\{\mathbf{s}(t), r(t), \mathbf{s}(t+1)\bigr\},\bigl\{\mathbf{o}_{k}(t), \mathbf{o}_{k}(t+1),u_{k}(t)\bigr\};
		\end{equation}
		
		\STATE Sample a random mini-batch of $b$ transitions from $\mathcal{B}$:
		\begin{equation}\nonumber
			\bigl\{\mathbf{s}(i), r(i), \mathbf{s}(i+1)\bigr\},\bigl\{\mathbf{o}_{k}(i), \mathbf{o}_{k}(i+1),u_{k}(i)\bigr\}, \forall k;
		\end{equation}
		
		\STATE Derive $K$ individual $Q$ values:
		\begin{equation}\nonumber
			\begin{aligned}
				\mathbf{Q}&=\bigl\{Q_{\bm \theta_{1}}\big(\mathbf{o}_{1}(i),u_{1}(i)\big),\dots,Q_{\bm \theta_{K}}\big(\mathbf{o}_{K}(i),u_{K}(i)\big)\bigr\},\\
				\mathbf{Q'}&=\bigl\{Q_{\bm \theta_{1}}\big(\mathbf{o}_{1}(i+1),u_{1}(i+1)\big),\dots,\\
				&\quad\quad\quad\quad\quad\quad\quad\quad Q_{\bm\theta_{K}}\big(\mathbf{o}_{K}(i+1),u_{K}(i+1)\big)\bigr\};
			\end{aligned}
		\end{equation}
		
		\STATE Compute weights for mixing network, then get $Q^{\boldsymbol{\pi}}_{\bm \sigma}\big(\mathbf{s}(i),\mathbf{u}(i)\big)$ and $Q^{\boldsymbol{\pi}}_{\bm \sigma}\big(\mathbf{s}(i+1),\mathbf{u}(i+1)\big)$:
		\begin{equation}\nonumber
			\begin{aligned}
				\bm \sigma(i)&\leftarrow\mathcal{H}_{\bm \delta}(\mathbf{s}(i)),\\
				\bm \sigma(i+1)&\leftarrow\mathcal{H}_{\bm \delta}\big(\mathbf{s}(i+1)\big),\\
				Q^{\boldsymbol{\pi}}_{\bm \sigma}\big(\mathbf{s}(i),\mathbf{u}(i)\big)&\leftarrow\mathcal{M}_{\bm \sigma(i)}(\mathbf{Q}),\\
				Q^{\boldsymbol{\pi}}_{\bm \sigma}\big(\mathbf{s}(i+1),\mathbf{u}(i+1)\big)&\leftarrow\mathcal{M}_{\bm \sigma(i+1)}(\mathbf{Q}');
			\end{aligned}
		\end{equation}
		
		\STATE Set $y(i) = r(i) + \gamma\displaystyle\max_{\mathbf{u}(i+1)}\bigl\{Q^{\boldsymbol{\pi}}_{\bm \sigma}\big(\mathbf{s}(i+1),\mathbf{u}(i+1)\big)\bigr\}$;
		\STATE Update all the agent networks and hyper-network by minimizing the following loss:
		\begin{equation}\nonumber
			L(\boldsymbol{\theta},\bm \sigma,\bm \delta)=\frac{1}{b}\sum_{i=1}^{b}\big[ y(i)-Q^{\boldsymbol{\pi}}_{\bm \sigma}\big(\mathbf{s}(i),\mathbf{u}(i)\big)\big]^{2}.
		\end{equation}
		\ENDFOR
	\end{algorithmic}
\end{algorithm}

\section{Simulation and Analysis}\label{section_results}
In this section, we will perform extensive computer simulation using the network simulator defined in Section \ref{section_smpf}. The simulated network includes many practical network mechanisms as shown in Appendix \ref{appedix_a}. Table~\ref{tb_67} summarizes the parameters employed in the simulation.

%\begin{table}[htp]
%\caption{BS Parameters used in simulation}\label{tb_6}
%\centering
%\begin{tabular}{ll}
%	\toprule[1.5pt]
%	Parameters                       & Values      \\ \midrule[1.5pt]
%	Transmit power for each RB       & 18 dBm      \\
%	Number of RB for each RBG        & 3           \\
%	Number of RBGs                   & 3           \\
%	Frequency bandwidth for each RBG & 10MHz       \\
%	Noise power density              & -174 dBm/Hz \\
%	Minimum MCS                      & 1           \\
%	Maximum MCS                      & 29          \\
%	Mximum number of HARQ            & 8           \\
%	Feedback period of HARQ          & 8           \\
%	Initial RB CQI value             & 4           \\
%	\bottomrule[1.5pt]
%\end{tabular}
%\end{table}

For illustration purposes, only three RBGs are deployed in the BS with each RBG composed of three resource blocks (RBs). In particular, each HARQ process can have at most eight re-transmissions with the ACK/NACK message is returned with a delay of seven TTIs.

\subsection{Model Training}
The agents are trained using the parameters listed in Table~\ref{tb_67}. Each epoch represents one scheduling process of $1000$~TTIs. In each epoch, BS will continue to accept new users while all agents perform RBG allocation until the end of the epoch. This process is designed to produce the experience data composed of the state-action pairs and the corresponding rewards. After that, each agent will select a mini-batch of experience data to update its parameters. $10$ batches of batch size of $256$ each are simulated.

%\begin{table}[htp]
%\caption{Parameters for training.}\label{tb_7}
%\centering
%\begin{tabular}{ll}
%	\toprule[1.5pt]
%	Parameters                      & Values   \\ \midrule[1.5pt]
%	Epoch				            & 100      \\
%	Duration of one experiment      & 1000~TTIs \\
%	Learning rate                   & 1e-3     \\
%	Learning rate decay             & 1e-7     \\
%	Batches                         & 10       \\
%	Batch size                      & 256      \\
%	Replayer capacity               & 2000     \\
%	$\epsilon$                      & 1e-2     \\
%	Initial number of users         & 5        \\
%	Average of Possion distribution ($\lambda$) & 1e-2     \\
%	\bottomrule[1.5pt]
%\end{tabular}
%\end{table}

\begin{table*}[]
	\caption{Parameters for simulation.}\label{tb_67}
	\centering
	\begin{tabular}{ll|ll}
		\hline
		\multicolumn{2}{l|}{BS Parameters used in simulation}        & \multicolumn{2}{l}{Parameters for training.}   \\ \hline
		Parameters                                  & Values         & Parameters                       & Values      \\ \hline
		Epoch                                       & 100            & Transmit power for each RB       & 18 dBm      \\
		Duration of one experiment                  & 1000$\sim$TTIs & Number of RB for each RBG        & 3           \\
		Learning rate                               & 1e-3           & Number of RBGs                   & 3           \\
		Learning rate decay                         & 1e-7           & Frequency bandwidth for each RBG & 10MHz       \\
		Batches                                     & 10             & Noise power density              & -174 dBm/Hz \\
		Batch size                                  & 256            & Minimum MCS                      & 1           \\
		Replayer capacity                           & 2000           & Maximum MCS                      & 29          \\
		$\epsilon$                                  & 1e-2           & Mximum number of HARQ            & 8           \\
		Initial number of users                     & 5              & Feedback period of HARQ          & 8           \\
		Average of Possion distribution ($\lambda$) & 1e-2           & Initial RB CQI value             & 4           \\ \hline
	\end{tabular}
\end{table*}

\begin{figure}[htp]
\centering
\includegraphics[width=\linewidth]{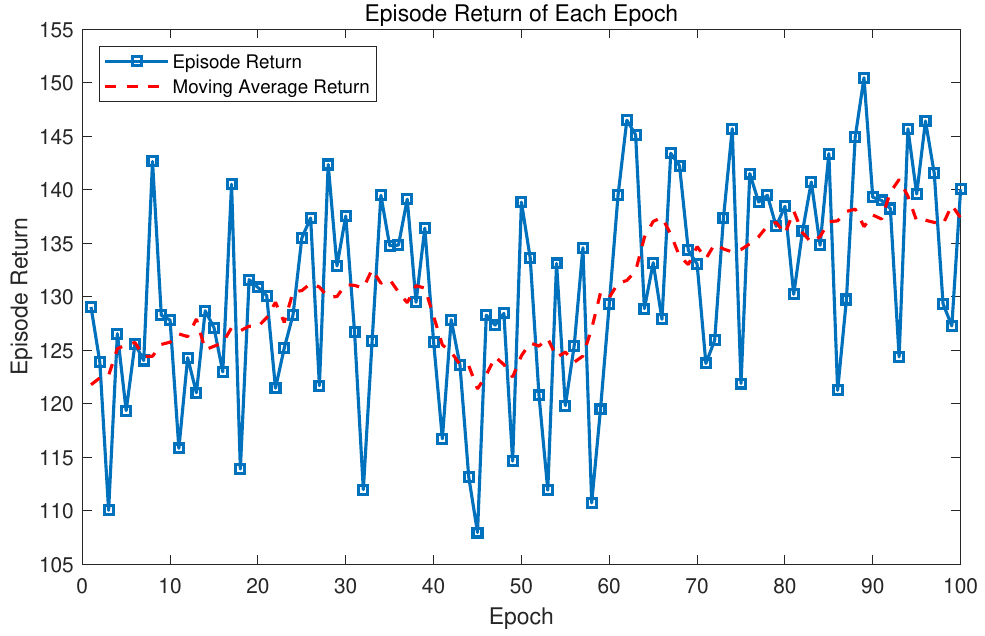}
\caption{Episode reward as a function of epochs.}
\label{fig_train_reward}
\end{figure}
As shown in Fig.~\ref{fig_train_reward}, the episode reward increased with the training epoch throughout the training process.

\subsection{Performance Comparison}
Next, we compare the performance of the proposed scheduler against the conventional OPS, RRS and GPFS, whose operating principle can be found in Appendix \ref{appedix_b}. It is well-known that the performance of GPFS is influenced by its two parameters, namely $\alpha_1$ and $\alpha_2$. For fair comparison, we will first select an appropriate $\alpha_1$ value to optimize the resulting 5TUDR performance using computer simulation.

\begin{figure}[h]
\centering
\includegraphics[width=1.1\linewidth]{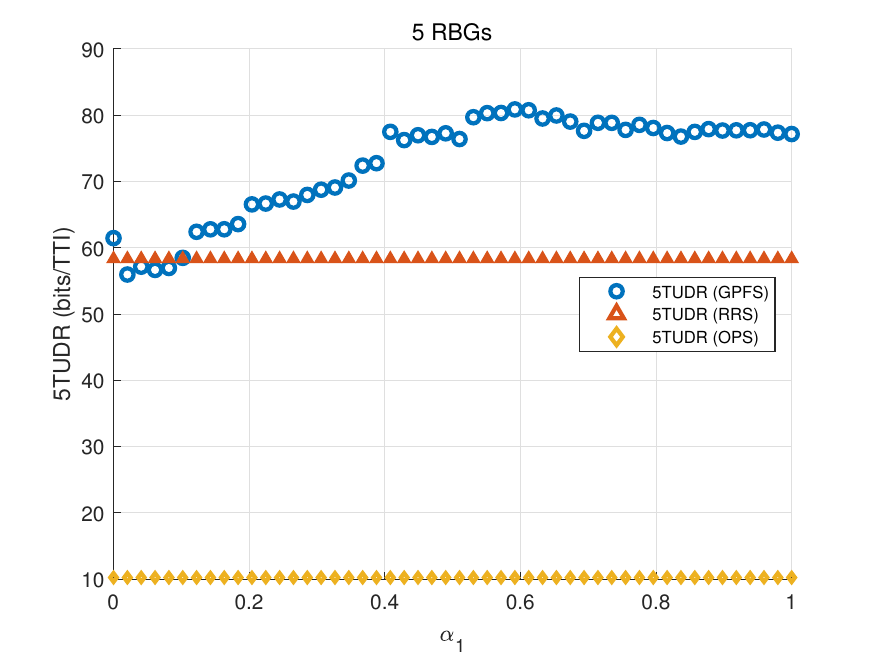}
\caption{The 5TUDR performance as a function of $\alpha_1$ with $\alpha_2=1$.}
\label{fig_sample_fairness}
\end{figure}

Fig.~\ref{fig_sample_fairness} shows the 5TUDR performance averaged over $1000$ experiments as a function of $\alpha_{1}$ for GPFS, RRS and OPS. Inspection of Fig.~\ref{fig_sample_fairness} reveals that the 5TUDR performance of GPFS initially increases with $\alpha_{1}$ before it saturates after $\alpha_{1}=0.5$. This is because that a larger $\alpha_{1}$ encourages GPFS to give higher scheduling priorities to users of better channel conditions, which initially improves both the AUDR and 5TUDR performance by more efficiently utilizing the network resources. However, any further increase in $\alpha_{1}$ deteriorated the data rates of the users of poor channel conditions, which became detrimental to the 5TUDR performance of the network. In contrast, the performance of  RRS and OPS is independent of $\alpha_{1}$ that is not a parameter in RRS and OPS. It is evidenced from Fig.~\ref{fig_sample_fairness} that OPS had the worst 5TUDR performance while GPFS with $\alpha_{1}=0.5$ outperformed OPS and RRS in terms of 5TUDR. In the sequel, GPFS with $\alpha_{1}=0$ and $\alpha_{2}=1$ is referred to as GPFS1 while GPFS with $\alpha_{1}=0.5$ and $\alpha_{2}=1$ as GPFS2.

\begin{figure}[h]
\centering
\includegraphics[width=1.1\linewidth]{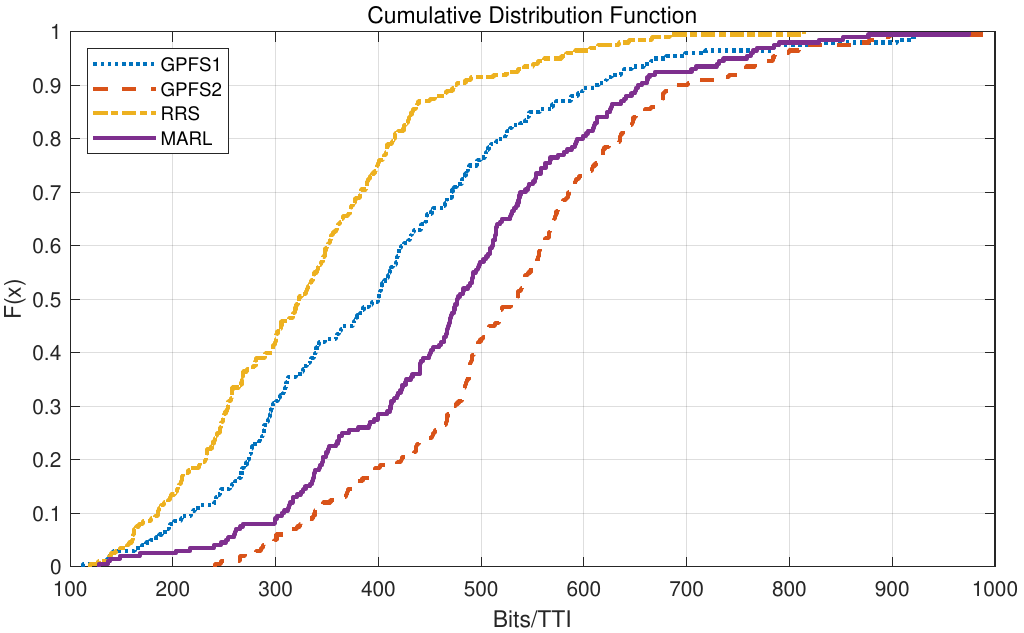}
\caption{CDF of the AUDR achieved by the proposed algorithm as compared to PFS and RRS.}
\label{fig_eval_avg}
\end{figure}

\begin{figure}[h]
\centering
\includegraphics[width=1.1\linewidth]{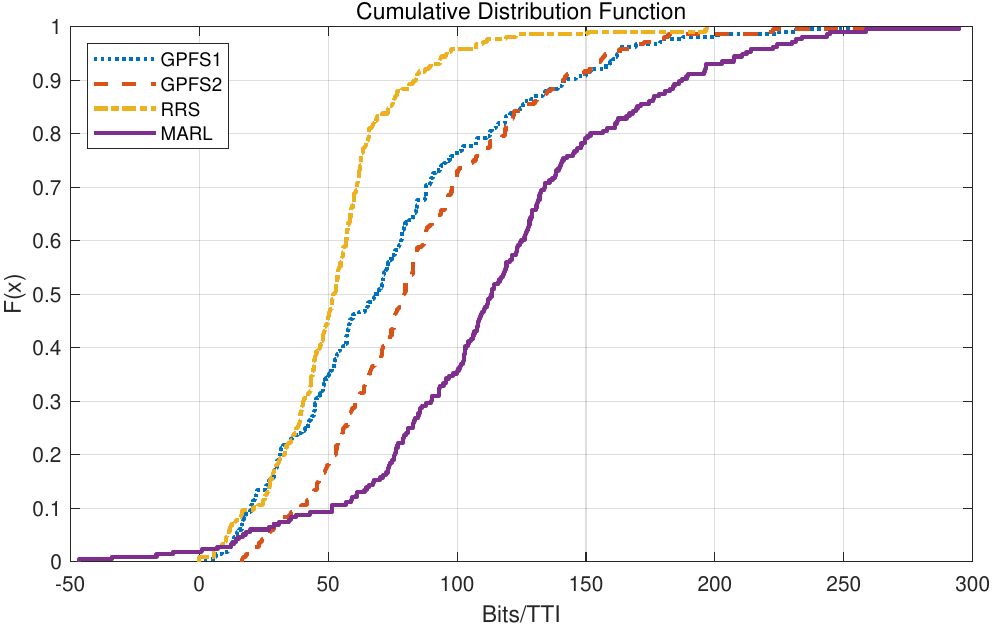}
\caption{CDF of the 5TUDR achieved by the proposed algorithm as compared to PFS and RRS.}
\label{fig_eval_fairness}
\end{figure}

Next, we will compare the AUDR and 5TUDR performance of the proposed MARL-based scheduler against GPFS1, GPFS2 and RRS. More specifically, we compute the improvement of the proposed scheduler over the three aforementioned conventional schedulers using $100$ experiments. Fig.~\ref{fig_eval_avg} and Fig.~\ref{fig_eval_fairness} plot the cumulative distribution function (CDF) of the achieved AUDR and 5TUDR, respectively. As shown in Fig.~\ref{fig_eval_avg}, the proposed scheduler substantially outperformed GPFS1 and RRS in terms of AUDR while suffering from marginal AUDR degradation as compared to GPFS2 that is optimized for AUDR as shown in Fig.~\ref{fig_sample_fairness}. More specifically, the median AUDR achieved by the proposed MARL-based scheduler was $456.94$~Bits/TTI as compared to $379.80$~Bits/TTI for GPFS1 and $322.0$~Bits/TTI for RRS, respectively. It is also observed in Fig.~\ref{fig_eval_avg} that GPFS2 had the highest median AUDR of $514.97$~Bits/TTI. Furthermore, Fig.~\ref{fig_eval_fairness} shows the CDF of the 5TUDR obtained. Clearly, the proposed scheduler significantly outperformed all three conventional schedulers in terms of 5TUDR, which is evidenced by the heavy tail of the CDF curve labelled as ``MARL". Specifically, the proposed scheduler achieved a median 5TUDR of $113.40$~Bit/TTI, which is a significant improvement as compared to GPFS2, GPFS1 and RRS. The median 5TUDR values obtained by GPFS2, GPFS1 and RRS are $79.56$~Bit/TTI,  $68.74$~Bit/TTI and $52.37$~Bit/TTI, respectively. Based on the discussions above, we can see that the proposed MARL-based scheduler is able to achieve substantially better 5TUDR performance (i.e. higher fairness) while maintaining a considerably large AUDR.

\subsection{Analysis of the Learned Policy}
To shed light on the policy learned by the proposed MARL-based scheduler, we will compare the scheduling preference of the proposed scheduler, GPFS1, GPFS2 and RRS. In the following simulation, we consider five users competing for three RBGs for bursty traffic over $500$~TTIs.

\subsubsection{More Transmission Time and Shorter Residence Time}
\begin{figure}[h]
\centering
\includegraphics[width=1.1\linewidth]{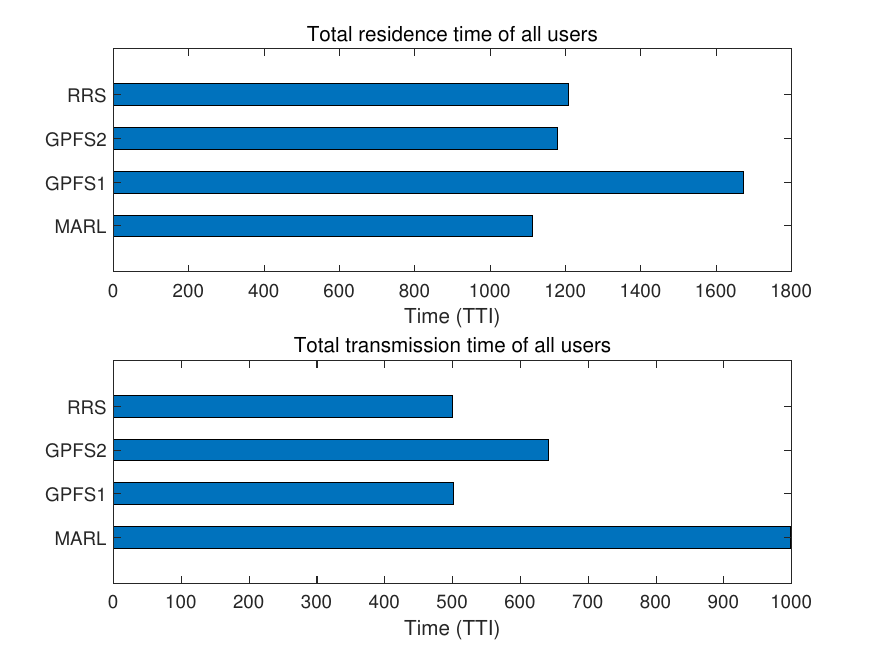}
\caption{Total residence time and total transmission time over all users.}
\label{fig_stay_alloc}
\end{figure}

We will first investigate two metrics, namely the total transmission time and the total residence time of each user in the system. The transmission time of a user counts the number of TTIs that the user is allocated with RBGs for transmission. For instance, if three RBGs are allocated to three different users in one TTI, then the transmission time is counted as three. In contrast, if all three RBGs are allocated to the same user in one TTI, then the corresponding transmission time becomes one. In other words, the transmission time is designed to measure how often users are served over a period of time. Furthermore, the residence time is defined as the number of TTIs that a user stays in the network before it completes its bursty transmission and leaves the network. Note that the residence time includes the time that a user spends on waiting for RBG allocation. We compare the total transmission time and the total residence time over all users for different schedulers in Fig.~\ref{fig_stay_alloc}. Inspection of Fig.~\ref{fig_stay_alloc} suggests that the proposed MARL-based scheduler resulted in significantly reduced total residence time, which indicates that the proposed scheduler can finish the requested transmissions for all users within a much shorter time period by more efficiently utilizing the same network resources. Furthermore, the total transmission time over all users was significantly increased, which implies that the proposed scheduler tends to divide the available RBGs to different users and subsequently, all users were scheduled more frequently. This also explains the shorter residence time of all users as all users had more opportunities for being served. The results shown in  Fig.~\ref{fig_stay_alloc} support our observation that the proposed scheduler resulted in higher AUDR and 5TUDR performance.

\subsubsection{Diverse RBG Allocation}
As suggested in Fig.~\ref{fig_stay_alloc}, the proposed MARL-based scheduler prefers allocating RBGs to diverse users, in lieu of concentrating all RBGs to only a few users. In this section, we will use data from two specific TTIs in a specific experiment to illustrate the different behaviors of the schedulers under consideration. Table~\ref{alloc_log_1} and Table~\ref{alloc_log_2} document the RBG allocation results in TTI=$1$ and TTI=$100$, respectively. In Table~\ref{alloc_log_1}, the first section details the counter information for each user. For the next four sections, we compare how each scheduler allocated each RBG. In particular, we use ``T" and ``F" to represent if the corresponding RBG was allocated or not allocated to the user indicated by the column index. For instance,  the proposed scheduler allocated ``RBG\#2" to ``UE\#1" while ``RBG\#1" and ``RBG\#3" to ``UE\#4" who has better RSRP and a smaller data package for transmission indicated by the ``Buffer" counter. In contrast, the three conventional schedulers allocated all RBGs to the same user as all users have the identical CQI value across all RBGs.

%\begin{table}[htp]
%	\caption{Allocation example 1 (TTI=1)}\label{alloc_log_1}
%	\centering
%	\begin{tabular}{lllllll}
%		\toprule[1.5pt]
%		& Attr.       & UE\#1 & UE\#2 & UE\#3 & UE\#4 & UE\#5 \\ \midrule[1.5pt]
%		& RSRP   &   -99  &   -90  &   -96  &  -88   &  -70   \\
%		& Buffer &  94768   &  96288   & 15528 &  4400   &   36400  \\
%		& HUDR    &  0.0  &  0.0   &  0.0   &  0.0   &  0.0   \\
%		& CQI\#1   &  4  &  4   &  4   &  4   &   4  \\
%		& CQI\#2   &  4  &  4   &  4   &  4   &   4  \\
%		& CQI\#3   &  4  &  4   &  4   &  4   &   4  \\ \midrule[1.0pt]
%		
%		& RBG\#1 &  F   &  F   &  F   &  T   &  F   \\
%		MARL & RBG\#2 &  T   &  F   &  F   &   F  &  F   \\
%		& RBG\#3 &  F   &  F   &  F   &  T  &  F  \\ \midrule[1.0pt]
%		
%		& RBG\#1 &  T   &  F   &  F   &  F   &  F   \\
%		GPFS1   & RBG\#2 &  T   &  F   &  F   &  F   &  F  \\
%		& RBG\#3 &  T   &  F   &  F   &  F   &  F  \\ \midrule[1.0pt]
%		
%		& RBG\#1 &  T   &  F   &  F   &  F   &  F   \\
%		GPFS2   & RBG\#2 &  T   &  F   &  F   &  F   &  F  \\
%		& RBG\#3 &  T   &  F   &  F   &  F   &  F  \\ \midrule[1.0pt]
%		
%		& RBG\#1 &  T   &  F   &  F   &  F   &  F  \\
%		RRS  & RBG\#2 &  T   &  F   &  F   &  F   &  F  \\
%		& RBG\#3 &  T   &  F   &  F   &  F   &  F  \\
%		\bottomrule[1.5pt]
%	\end{tabular}
%\end{table}

\begin{table*}[]
	\caption{Allocation example 1 (TTI=1)}\label{alloc_log_1}
	\centering
	\begin{tabular}{|l|llllll|l|llllll|}
		\hline
		& Attr.  & UE\#1 & UE\#2 & UE\#3 & UE\#4 & UE\#5 &       & Attr.  & UE\#1 & UE\#2 & UE\#3 & UE\#4 & UE\#5 \\ \hline
		& RSRP   & -99   & -90   & -96   & -88   & -70   &       & RSRP   & -99   & -90   & -96   & -88   & -70   \\
		& Buffer & 94768 & 96288 & 15528 & 4400  & 36400 &       & Buffer & 94768 & 96288 & 15528 & 4400  & 36400 \\
		& HUDR   & 0.0   & 0.0   & 0.0   & 0.0   & 0.0   &       & HUDR   & 0.0   & 0.0   & 0.0   & 0.0   & 0.0   \\
		& CQI\#1 & 4     & 4     & 4     & 4     & 4     &       & CQI\#1 & 4     & 4     & 4     & 4     & 4     \\
		& CQI\#2 & 4     & 4     & 4     & 4     & 4     &       & CQI\#2 & 4     & 4     & 4     & 4     & 4     \\
		& CQI\#3 & 4     & 4     & 4     & 4     & 4     &       & CQI\#3 & 4     & 4     & 4     & 4     & 4     \\ \hline
		& RBG\#1 & F     & F     & F     & \textbf{T}     & F     &       & RBG\#1 & \textbf{T}     & F     & F     & F     & F     \\
		MARL  & RBG\#2 & \textbf{T}     & F     & F     & F     & F     & GPFS2 & RBG\#2 & \textbf{T}     & F     & F     & F     & F     \\
		& RBG\#3 & F     & F     & F     & \textbf{T}     & F     &       & RBG\#3 & \textbf{T}     & F     & F     & F     & F     \\ \hline
		& RBG\#1 & \textbf{T}     & F     & F     & F     & F     &       & RBG\#1 & \textbf{T}     & F     & F     & F     & F     \\
		GPFS1 & RBG\#2 & \textbf{T}     & F     & F     & F     & F     & RRS   & RBG\#2 & \textbf{T}     & F     & F     & F     & F     \\
		& RBG\#3 & \textbf{T}     & F     & F     & F     & F     &       & RBG\#3 & \textbf{T}     & F     & F     & F     & F     \\ \hline
	\end{tabular}
\end{table*}

After the network ran for $100$~TTIs, we investigated the network counter information as well as the scheduling results in Table~\ref{alloc_log_2}. For the proposed scheduler, ``UE\#3", ``UE\#4" and ``UE\#5" have already finished their transmission and left the network, which is indicated by "N/A" in Table~\ref{alloc_log_2}. As a result, only ``UE\#1" and ``UE\#2" were allocated RBGs. Despite the fact that ``UE\#2" had better channel conditions (indicated by its CQI and RSRP values) and a lower HUDR, the proposed scheduler was actually in favor of ``UE\#1". This scheduling decision could be explained by the fact that ``UE\#1" has a smaller buffer size and may be able to finish its transmission in a shorter time. In contrast, GPFS2 and RRS only finished the transmission for ``UE\#4" and ``UE\#5" by TTI=$100$ while GPFS1 had four users remaining in the network. Note that GPFS1 and  GPFS2 allocated more RBGs to ``UE\#1" as ``UE\#1" has the lowest HUDR among all remaining users.

\begin{table*}[htp]
	\caption{Allocation example 2 (TTI=100)}\label{alloc_log_2}
	\centering
	\begin{tabular}{|l|llllll|l|llllll|}
		\hline
		& Attr.  & UE 1 & UE 2 & UE 3 & UE 4 & UE 5 &       & Attr.  & UE 1 & UE 2 & UE 3 & UE 4 & UE 5 \\\hline
		& RSRP   & -99   & -90   & N/A   & N/A   & N/A   &       & RSRP   & -99   & -90   & -96   & N/A   & N/A   \\
		& Buffer & 86119 & 89886 & N/A   & N/A   & N/A   &       & Buffer & 84663 & 86246 & 5884  & N/A   & N/A   \\
		& HUDR   & 72.1  & 42.9  & N/A   & N/A   & N/A   &       & HUDR   & 81.6  & 182.3 & 132.1 & N/A   & N/A   \\
		& CQI\#1 & 7     & 9     & N/A   & N/A   & N/A   &       & CQI\#1 & 7     & 9     & 7     & N/A   & N/A   \\
		MARL  & CQI\#2 & 7     & 9     & N/A   & N/A   & N/A   & GPFS2 & CQI\#2 & 7     & 9     & 7     & N/A   & N/A   \\
		& CQI\#3 & 3     & 14    & N/A   & N/A   & N/A   &       & CQI\#3 & 3     & 14    & 9     & N/A   & N/A   \\ \cmidrule(){2-7} \cmidrule(){9-14}
		& RBG\#1 & \textbf{T}     & F     & F     & F     & F     &       & RBG\#1 & \textbf{T}     & F     & F     & F     & F     \\
		& RBG\#2 & \textbf{T}     & F     & F     & F     & F     &       & RBG\#2 & \textbf{T}     & F     & F     & F     & F     \\
		& RBG\#3 & F     & \textbf{T}     & F     & F     & F     &       & RBG\#3 & F     & \textbf{T}     & F     & F     & F     \\ \hline
		& RSRP   & -99   & -90   & -96   & N/A   & -70   &       & RSRP   & -99   & -90   & -96   & N/A   & N/A   \\
		& Buffer & 84563 & 87882 & 5360  & N/A   & 2433  &       & Buffer & 88587 & 82716 & 5828  & N/A   & N/A   \\
		& HUDR   & 82.8  & 158.6 & 139.3 & N/A   & 849.2 &       & HUDR   & 55.9  & 125.3 & 90.3  & N/A   & N/A   \\
		& CQI\#1 & 7     & 9     & 7     & N/A   & 25    &       & CQI\#1 & 7     & 9     & 7     & N/A   & N/A   \\
		GPFS1 & CQI\#2 & 7     & 9     & 7     & N/A   & 25    & RRS   & CQI\#2 & 7     & 9     & 7     & N/A   & N/A   \\
		& CQI\#3 & 3     & 14    & 9     & N/A   & 27    &       & CQI\#3 & 3     & 14    & 9     & N/A   & N/A   \\ \cmidrule(){2-7} \cmidrule(){9-14}
		& RBG\#1 & \textbf{T}     & F     & F     & F     & F     &       & RBG\#1 & \textbf{T}     & F     & F     & F     & F     \\
		& RBG\#2 & \textbf{T}     & F     & F     & F     & F     &       & RBG\#2 & \textbf{T}     & F     & F     & F     & F     \\
		& RBG\#3 & \textbf{T}     & F     & F     & F     & F     &       & RBG\#3 & \textbf{T}     & F     & F     & F     & F     \\ \hline
	\end{tabular}
\end{table*}

\begin{figure}[htp]
\centering
\includegraphics[width=0.8\linewidth]{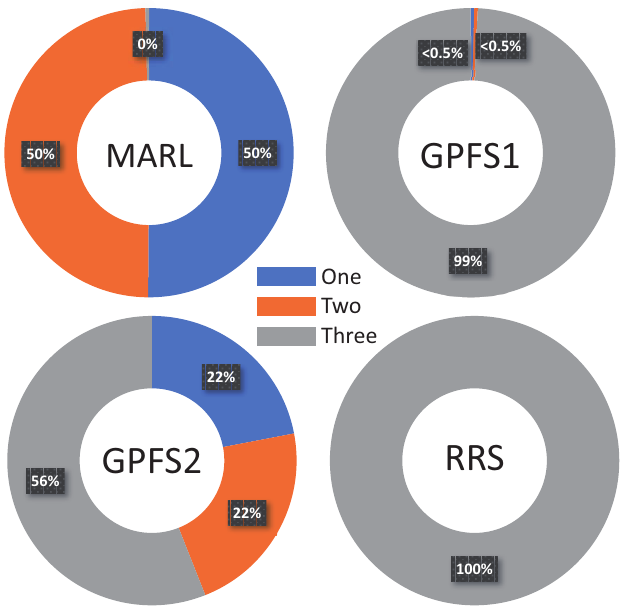}
\caption{Percentages of scheduled users who were allocated with ``One", ``Two" or ``Three” RBGs.}
\label{fig_dis_alloc}
\end{figure}

Finally, we analyze the number of RBGs that each scheduler allocates to a single user in the same TTI. Fig.~\ref{fig_dis_alloc} shows the percentages of scheduled users who were allocated with ``One", ``Two" or ``Three” RBGs in the same TTI, recalling that our simulation only considered three RBGs. As indicated in Fig.~\ref{fig_dis_alloc}, GPFS1 and RRS allocated all the RBGs to a single user almost throughout the entire simulation. In contrast, the proposed scheduler and GPFS2 chose to distribute their RBGs to different users more evenly. As a result, users are given more opportunities to access RBGs, which led to higher user fairness. In particular, the proposed scheduler never allocated all three RBGs to one single user, which makes our proposed scheduler very distinctive from the three conventional schedulers.

\subsubsection{Scheduling Preference}
\begin{figure}[htp]
\centering
\includegraphics[width=1.1\linewidth]{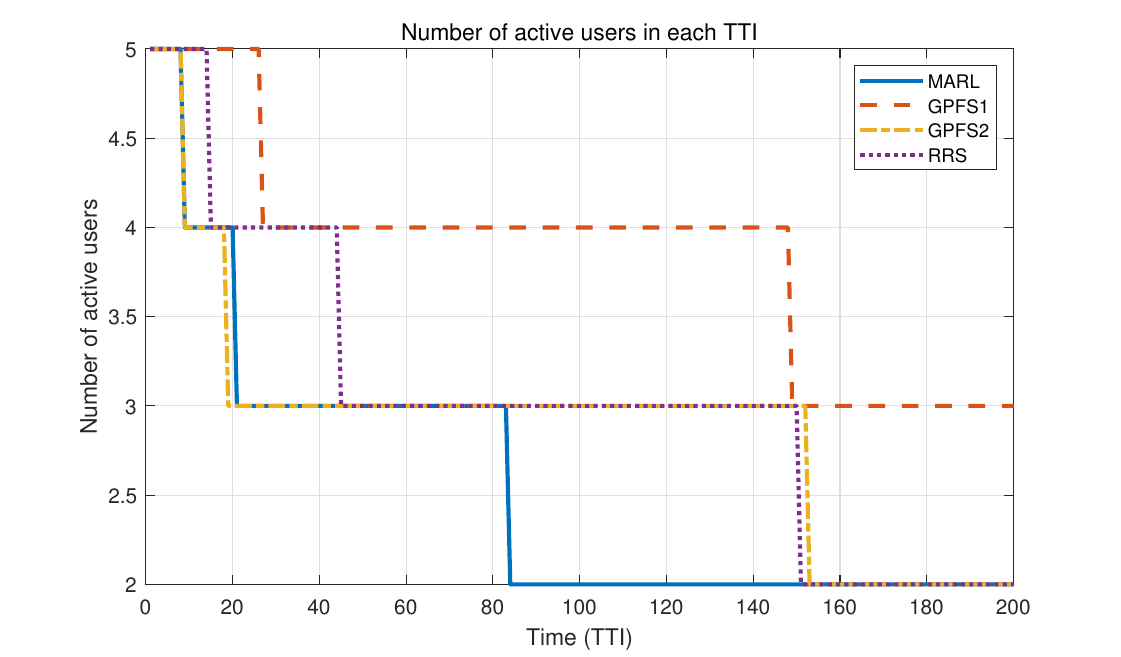}
\caption{Number of active users as a function of TTI.}
\label{fig_aum}
\end{figure}

Finally, we explore the scheduling preference of the proposed MARL-based scheduler. Unfortunately, the policy of the proposed scheduler is represented by a set of DNN parameters, which makes the analysis challenging. To cope with this challenge, we first analyze the scheduling log data as shown in Table~\ref{alloc_log_1} and Table~\ref{alloc_log_2}. Specifically, ``UE\#1" and ``UE\#2" in Table~\ref{alloc_log_1} and Table~\ref{alloc_log_2} had a much larger buffer size than the other three users. By TTI=$100$, the proposed scheduler has completed the data transmission for those users of a smaller buffer size without holding these users waiting in the network. As a result, the average user data rates for those users were improved, which in turns contributed to the increase in AUDR. Therefore, the results hint that the proposed scheduler prefers scheduling the users of a smaller buffer size first. To validate this insight, we first investigate the number of active users waiting for services in each TTI by different schedulers. Fig.~\ref{fig_aum} shows that the number of active users in the network was reduced from five to two only after $83$ TTIs by the proposed scheduler, meaning that three users finished their transmission within the first $83$ TTIs. In contrast, it took almost $150$ TTIs for RRS and GPFS1 to finish the transmission for three users while GPFS1 was not able to complete the transmission for all five users even after $500$ TTIs.

\begin{table}[htp]
\centering
\caption{Total RBG number allocated to users of a smaller buffer size during the first $100$~TTIs.}
\begin{tabular}{ccccc}
	\toprule[1.5pt]
	& MARL & GPFS1 & GPFS2 & RRS \\
	\midrule[1.5pt]
	UE\#3 & 127  & 75    & 73    & 84  \\
	UE\#4 & 16   & 5     & 7     & 15  \\
	UE\#5 & 24   & 25    & 20    & 30 \\
	\midrule[1.5pt]
	Sum   & 167  & 105   & 100   & 129\\
	\bottomrule[1.5pt]
\end{tabular}	\label{fig_alloc_to_low_buffer}
\end{table}

Furthermore, we examine the total number of RBGs assigned to ``UE\#3", ``UE\#4" and ``UE\#5" whose buffer sizes are $15528$~bits, $4400$~bits and $36400$~bits, respectively. Note that the RSRP values for ``UE\#3", ``UE\#4" and ``UE\#5" are $-96$~dB, $-88$~dB and $-70$~dB, respectively. Table~\ref{fig_alloc_to_low_buffer} shows the total RBG number allocated to each of these three users during the first $100$~TTIs, i.e. $300$~RBGs in total. We first observed that the proposed scheduler allocated more than $50\%$, i.e. $167$ out of the total $300$ RBGs to these three users of smaller buffer sizes. In addition, inspection of the simulation data revealed that it took only $9$~TTIs or $16$~RBGs for the proposed MARL-based scheduler to finish the transmission for ``UE\#4" of relatively high RSRP. Furthermore, the proposed scheduler allocated noticeably more RBGs to expedite the data transmission for ``UE\#3". As a result, the proposed scheduler was able to complete the transmission for ``UE\#3" after $83$ TTIs, which enabled the network to concentrate on serving the remaining users. By providing scheduling priority to the users of smaller buffer sizes, the proposed scheduler can improve the UDR of those users and subsequently, the AUDR and 5TUDR performance.

\begin{figure}[h]
\centering
\includegraphics[width=\linewidth]{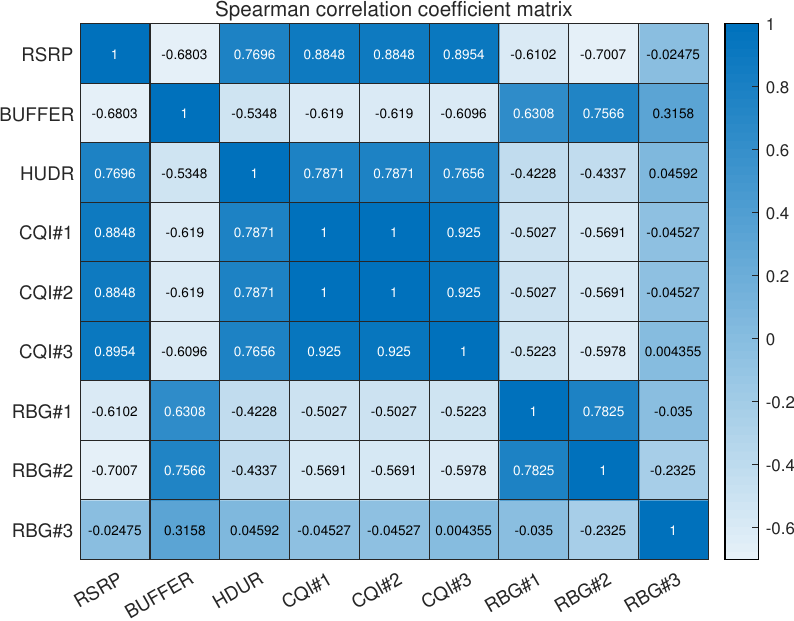}
\caption{Spearman correlation coefficient matrix.}
\label{fig_rl_corr}
\end{figure}

To better characterize the proposed scheduler, we calculate the Spearman correlation coefficient matrix \cite{sedgwick2014spearman} illustrated in Fig.~\ref{fig_rl_corr}. For instance, the correlation coefficient between the allocation result of ``RBG\#1" and buffer is $0.6308$, which implies that the allocation of ``RBG\#1" has a strong connection with the buffer size of the scheduled user.

In summary, the scheduling policy of the proposed MARL-based scheduler has the following characteristics. First, it prefers to distribute RBGs to different users in the same TTI. As a result, all users have more opportunities to access network resources for transmission, which effectively improves the user fairness. In addition, such a scheduling policy helps increase the average user data rate, which in turns increases the 5TUDR performance. Furthermore, the proposed scheduler is in favor of finishing the transmission requests from those users of a smaller buffer size. This feature indeed helps reduce the total time that those users spend in the network and subsequently, improve the user data rate for those users. Finally, despite that the scheduling policy of the proposed scheduler cannot be mathematically analyzed, some insights can be observed through the Spearman correlation coefficient matrix to identify the strongly correlated input-output factor pairs.

\section{CONCLUSION}\label{section_conclusion}
In this work, we have investigated the problem of RBG scheduling for users of bursty traffic in wireless networks. To provide fair opportunities to all users to access the available RBGs, a fairness-oriented scheduler has been formulated as a stochastic game by incorporating information from multiple network layers such as CQI and buffer size. In particular, a reward function has been devised to maximize the $5$\%-tile user data rate. Furthermore, a MARL-based learning approach has been proposed to learn the optimal scheduling policy by dividing the large state-space into multiple sub-problems each of which is handled by one agent. Extensive computer simulation has been performed to compare the performance of the proposed scheduler against the conventional GPFS, OPS and RRS. Simulation results have confirmed that the proposed scheduler can achieve impressive 5TUDR performance while maintaining good AUDR performance. Finally, we have investigated the characteristics of the scheduling policy learned by the proposed scheduler and found that the proposed scheduler distributes RBGs to more users in each TTI with higher preference towards users of a smaller buffer size.

There are several extensions of this study that can be further explored. One possible drawback of the proposed scheme is that the mathematical relationship between OP1 and OP2 has not been rigorously established. In addition, the optimality of the QMIX solution deserves further investigation. Finally, it is of great practical interest to compare the performance provided by different G's fairness index functions in the proposed reward function.

\section{Appendix: Network mechanism}\label{appedix_a}
In this appendix, we review some of the network mechanisms implemented in our network simulator that was used in generating the training and test data sets for our experiments.
\subsection{Out Loop Link Adaptation (OLLA)} Each UE measures the received SNR on each RB, and periodically reports the corresponding CQI values to the BS. The BS selects the appropriate Modulation and Coding Scheme (MCS) based on the received CQI. However, the reported CQI is distorted by noise. As a result, the MCS chosen based on the reported CQI is not always optimal. To compensate for the discrepancy between the chosen MCS and the optimal MCS, an OLLA process is carried out on the BS by adding a small offset $\alpha$ to the current CQI value $q$, i.e.
\begin{equation}
\bar{q}=\left[q+\alpha\right],
\end{equation}
where $[\cdot]$ is the rounding operator and $\bar{q}$ is the adjusted CQI for the MCS selection. In addition, $\alpha$ is updated every time when an ACK/NACK is received according to
\begin{equation}
\alpha=
\begin{cases}
\alpha+s_A, & \text{if an ACK is received}, \\
\alpha+s_N, & \text{if a NACK is received},
\end{cases}
\end{equation}
where $s_A>0$ and $s_N<0$ are the update rates.

\subsection{Transmission Block Formation}
After the BS decides the RBG allocation, it will calculate the transmission block (TB) size for each scheduled user. Specifically, for each scheduled user, the BS first computes the average adjusted CQI over the set of all RBs allocated to that user, and then use the adjusted CQI to select a proper MCS index. The chosen MCS index is then mapped to a spectral efficiency according to some operator-specific tables. Finally, the spectral efficiency and the number of allocated RBGs together decide the transmission block size.

Let $\mathcal{F}(\bar{q})$ denote the mapping from the CQI $\bar{q}$ to its spectral efficiencies (SE), and $\mathcal{I}$ a set of CQI levels over a specific group of RBs observed by a user. The maximum number of bits that can be loaded to this group of RBs by the user is given by
\begin{equation}
f(\mathcal{I})=|\mathcal{I}|\cdot
\mathcal{F} \left( \bigg\lfloor \frac{1}{|\mathcal{I}|}
\sum_{i}\mathcal{I}(i)
\bigg\rfloor \right),
\end{equation}
where $\lfloor\cdot\rfloor$ is the floor function. For instance, the estimated data rate of the  $n$-th user in the $k$-th RBG can be obtained by $R_{n,k}=f(\mathcal{I}_{n}(k))$, where $\mathcal{I}_{n}(k)$ is the set of all RBs in the $k$-th RBG measured by the $n$-th user. If multiple RBGs are allocated to the same user,  the same MCS is utilized to convey one transport block (TB) whose size is given by $T_{n}=f(\mathcal{I}_{n})$, where $\mathcal{I}_{n}$ is the set of all RBs allocated to the $n$-th user.

\subsection{HARQ and Retransmission}
When a TB is formed, its data will be loaded into a HARQ buffer and stored there until the transmission completes (i.e., receive an ACK) or the data is dropped due to five consecutive transmission failures (i.e., five NACKs). The BS arranges at most eight HARQ processes (each with a HARQ buffer) for each active user at a time.
When a NACK message is received, a retransmission is triggered and the RBGs initially assigned to the failed transmission are reserved for retransmission. The MCS chosen for the retransmission remains the same to ensure that the same TB can be reloaded to the delegated RBGs again.

\subsection{Conventional Scheduling Schemes for Benchmarking}\label{appedix_b}
In this Appendix, we provide some definitions of GPFS, OPS and RRS.
\subsubsection{Generalized Proportional Fairness Scheduling (GPFS)}
In GPFS, each RBG is independently scheduled according to the Proportional Fairness (PF) values of all active users. More specifically, the PF value of the $n$-th user on the $k$-th RBG at the $t$-th TTI is defined as
\begin{equation}
\zeta^{n}_{k}(t)=\frac{(R^{n}_{k}(t))^{\alpha_1}}{(\hat{T}^n(t))^{\alpha_2}},
\label{eq1}
\end{equation}
where $R^{n}_{k}(t)$ is the achievable data rate of the $n$-th user on the $k$-th RBG at the $t$-th TTI while $\hat{T}^n(t)$ is the user's moving average throughput expressed as
\begin{equation}
\hat{T}^n(t)=\left(1-\chi\right)\hat{T}^n(t-1)+\chi T^n(t-1).
\label{eq:2}
\end{equation}
with $\chi$ and $T^n(t-1)$ being the moving average coefficient and the actual TB size of the $n$-th user at the ($t-1$)-th TTI. Furthermore, $\alpha_1$ and $\alpha_2$ are two design parameters within $[0,1]$. For $\alpha_1=0$ and $\alpha_2=1$, GPFS concentrates on giving priorities to users of low average data rate in the past. In contrast, $\alpha_1=1$ and $\alpha_2=0$ degenerates GPFS to the conventional OPS.

Users of larger PF values are given higher priorities to be scheduled for the RBG under consideration.
\begin{equation}
P_{GPFS}^*(k)=\underset{n}{\rm argmax} \{\zeta^{n}_{k}\}.
\end{equation}
Note that the PF value of the same user may vary across different RBGs in the same TTI.

\subsubsection{Opportunistic Scheduling (OPS)} The opportunistic scheduling allocates an RBG to the user who can achieve the highest estimated data rate and takes the following form:
\begin{equation}
P_{OP}^*(k)=\underset{n}{\rm argmax}\{R^{n}_{k}\}.
\label{eq:oppo}
\end{equation}
For full buffer traffic, OPS is considered the optimal algorithm for achieving the highest network throughput and AUDR. However, this conclusion is not necessarily true for bursty traffic.

\subsubsection{Round Robin Fashion Scheduling (RRS)} RRS is the classic scheduling algorithm that allocates all available RBGs to one user at a time and serves all users in turns. Note that new users are appended to the end of the queue for scheduling if bursty traffic is considered.

\section*{Acknowledgment}
The authors would like to thank the editors and reviewers for their constructive comments, which helped to substantially improve the quality of this article.

\section*{Financial Support}
This work was supported by National Key Research and Development Program of China (2020YFB1807700).

%\bibliographystyle{IEEEtran}
%\bibliography{reference}


\end{document}